\documentclass[3p,twocolumn,10pt,number]{elsarticle}

\usepackage{graphicx}
\usepackage[english]{babel}
\usepackage{xspace}
\usepackage{siunitx} 

\usepackage{amssymb}
\usepackage{amssymb}
\usepackage{amsthm}
\usepackage{amsfonts}
\usepackage{amsopn}
\usepackage{amsmath}

\usepackage{color}
\usepackage{csquotes}
\usepackage{hyperref}

\newcommand{\omegan}{\ensuremath{\omega_{\mathrm{n}}}\xspace}
\newcommand{\omegaHg}{\ensuremath{\omega_{\mathrm{Hg}}}\xspace}

\newcommand{\dn}{\ensuremath{d_\mathrm{n}}\xspace}
\newcommand{\gamman}{\ensuremath{\gamma_{\mathrm{n}}}\xspace}
\newcommand{\mun}{\ensuremath{\mu_{\mathrm{n}}}\xspace}
\newcommand{\gammaHg}{\ensuremath{\gamma_\mathrm{Hg}}\xspace}
\newcommand{\dHg}{\ensuremath{d_{\magHg}}\xspace}

\newcommand{\ecm}{\si{\elementarycharge\ensuremath{\cdot}\centi\meter}\xspace}
\newcommand{\fT}{\si{\femto\tesla\xspace}}

\newcommand{\Evec}{\ensuremath{\mathbf{E}}\xspace}
\newcommand{\Bvec}{\ensuremath{\mathbf{B_0}}\xspace}

\newcommand{\Bz}{\ensuremath{B_\mathrm{0}}\xspace}

\newcommand{\magHg}{\ensuremath{{}^{199}\text{Hg}}\xspace}
\newcommand{\lampHg}{\ensuremath{{}^{204}\text{Hg}}\xspace}

\DeclareMathOperator{\arcsinh}{arcsinh}

\newcommand\solidrule[1][0.6cm]{\rule[0.5ex]{#1}{1.2pt}}
\newcommand\dashedrule{\mbox{%
  \solidrule[2mm]\hspace{2mm}\solidrule[2mm]\hspace{2mm}\solidrule[2mm]}}

\newcommand{\OpenCircleBlue}{\ensuremath{\textcolor{blue}{\circ}}\xspace}

\newcommand{\OpenDiamondBlack}{\ensuremath{\textcolor{black}{\Diamond}}\xspace}

\newcommand{\FilledCircleRed}{\ensuremath{\textcolor{red}{\bullet}}\xspace}

\newcommand{\FilledSquareRed}{\ensuremath{\textcolor[rgb]{1,0,0}{\blacksquare}}\xspace}
\newcommand{\PlotLineBrown}{\textcolor[rgb]{0.6,0.4,0.2}{\solidrule}\xspace}

\newcommand{\PlotLineRed}{\textcolor[rgb]{1,0,0}{\solidrule}\xspace}
\newcommand{\PlotLineBlack}{\textcolor{black}{\solidrule}\xspace}

\newcommand{\PlotLineYellowDashed}{\textcolor{yellow}{\dashedrule}\xspace}

\newcommand{\PlotLineBlackDashed}{\textcolor{black}{\dashedrule}\xspace}

\journal{Nuclear Instruments and Methods in Physics Research A}

\begin{document}

\begin{frontmatter}

\title{Demonstration of sensitivity increase in mercury free-spin-precession magnetometers due to laser-based readout for neutron electric dipole moment searches}

\author[LPC]{G.~Ban}
\author[PSI]{G.~Bison\corref{cor1}}
\ead{georg.bison@psi.ch}
\author[JAG]{K.~Bodek}
\author[PSI]{M.~Daum}
\author[PSI,ETH]{M.~Fertl\corref{cor2}\fnref{fn1}}
\ead{mfertl@uw.edu}
\author[PSI,ETH]{B.~Franke\fnref{fn2} }
\author[FRAP]{Z.~D.~Gruji\'c}
\author[JGU]{W.~Heil}
\author[PSI]{M.~Horras}
\author[FRAP]{M.~Kasprzak\fnref{fn3}}
\author[LPSC]{Y.~Kermaidic\fnref{fn4}}
\author[PSI,ETH]{K.~Kirch}
\author[FRAP,JGU]{H.-C.~Koch\fnref{fn3}}
\author[PSI,ETH]{S.~Komposch}
\author[HNI]{A.~Kozela}
\author[ETH]{J.~Krempel}
\author[PSI]{B.~Lauss}
\author[LPC]{T.~Lefort}
\author[PSI]{A.~Mtchedlishvili}
\author[LPSC]{G.~Pignol}
\author[ETH,UBE]{F.M.~Piegsa}
\author[KLU,PSI,ETH]{P.~Prashanth}
\author[LPC]{G.~Qu\'em\'ener}
\author[JAG,ETH]{M.~Rawlik}
\author[LPSC]{D.~Rebreyend}
\author[PSI,ETH]{D.~Ries\fnref{fn7}}
\author[CSN]{S.~Roccia}
\author[JAG]{D.~Rozpedzik}
\author[PSI]{P.~Schmidt-Wellenburg}
\author[KUL]{N.~Severijns}
\author[FRAP]{A.~Weis}
\author[JAG,ETH]{G.~Wyszynski}
\author[JAG]{J.~Zejma}
\author[PSI]{G.~Zsigmond}

\address[LPC]{Laboratoire de Physique Corpusculaire, Caen, France}
\address[PSI]{Paul Scherrer Institute, Villigen, Switzerland}
\address[JAG]{M.~Smoluchowski Institute of Physics, Jagiellonian University, Cracow, Poland}
\address[ETH]{ETH Z\"urich, Institute for Particle Physics and Astrophysics, Z\"urich, Switzerland}
\address[FRAP]{University of Fribourg, Switzerland}
\address[LPSC]{Laboratoire de Physique Subatomique et de Cosmologie, Grenoble, France}
\address[JGU]{Institut f\"ur Physik, Johannes-Gutenberg-Universit\"at, Mainz, Germany}
\address[HNI]{Henryk Niedwodnicza\'nski Institute for Nuclear Physics, Cracow, Poland}
\address[UBE] {Laboratory for High Energy Physics, Albert Einstein Center for Fundamental Physics, University of Bern, Switzerland}
\address[KUL]{Katholieke Universiteit, Leuven, Belgium}
\address[CSN]{Centre de Sciences Nucl\`eaires et de Sciences de la Mati\`ere, Orsay, France}

\fntext[fn1]{Now at University of Washington, Seattle, USA}
\fntext[fn2]{Now at TRIUMF, Vancouver, Canada}
\fntext[fn3]{Now at Paul Scherrer Institut, Villigen, Switzerland}
\fntext[fn4]{Now at Max-Planck-Institut f\"ur Kernphysik, Heidelberg, Germany}
\fntext[fn7]{Now at Institute for Nuclear Chemistry, Johannes-Gutenberg-Universit\"at, Mainz, Germany}

\cortext[cor1]{Corresponding author}
\cortext[cor2]{Corresponding author}

\begin{abstract}

We report on a laser based \magHg co-magnetometer deployed in an experiment searching for a permanent electric dipole moment of the neutron.
We demonstrate a more than five times increased signal to-noise-ratio in a direct comparison measurement with its \lampHg discharge bulb-based predecessor.
An improved data model for the extraction of important system parameters such as the degrees of absorption and polarization is derived.
Laser- and lamp-based data-sets can be consistently described by the improved model which permits to compare measurements using the two different light sources and to explain the increase in magnetometer performance.
The laser-based magnetometer satisfies the magnetic field sensitivity requirements for the next generation nEDM experiments.

\end{abstract}


\end{frontmatter}



\section{\label{sec:Intro}Introduction}
The search for a permanent electric dipole moment of the neutron (nEDM, \dn) is considered one of the most important experiments in low-energy particle physics \cite{Raidal2008}. 
Any permanent electric dipole moment (EDM) violates parity P and time-reversal symmetry T. 
Invoking the CPT theorem (C charge conjugation symmetry), an EDM also violates CP symmetry, an ingredient needed to explain the observed matter-antimatter asymmetry of the Universe \cite{Sakharov1967}. This motivates EDM searches in molecular or atomic systems, nucleons, and bare leptons. A detailed review of the sensitivity provided by the different systems to the underlying physics can be found in \cite{Engel2013}. The neutron as bare nucleon offers direct access to its EDM without being embedded in a more complex electronic or nuclear structure.

The latest experimental limit  $\dn \leq \SI{3E-26}{\ecm} \,(\SI{90}{\percent}\,\mathrm{CL})$ \cite{Pendlebury2015PRD}, a reanalysis of the data of Ref.\ \cite{Baker2006}, constrains the theta parameter in quantum chromodynamics (QCD) to $\theta_{\mathrm{QCD}}\leq10^{-10}$ and rules out many models for physics beyond the Standard Model \cite{McKeen2012}.

A widely used approach in the search for \dn is to measure the spin precession (Larmor) frequency of confined ultra-cold neutrons (UCN) exposed to a superposition of an electric \Evec and magnetic \Bvec field configuration by applying Ramsey's method of separated oscillatory fields \cite{Ramsey1949}.
A non-vanishing \dn leads to a linear dependence of the neutron spin precession frequency \omegan on $|\Evec|$
\begin{equation}
\omegan = \frac{2\, \mun |\Bvec|}{\hbar} \pm \frac{2\, \dn |\Evec|}{\hbar} = \gamma_\mathrm{n} |\Bvec| \pm \Delta\omega_{\mathrm{EDM}}(|\Evec|)
\label{eqn:omegan}
\end{equation}
where \mun is  the magnetic dipole moment of the neutron, $\gamma_\mathrm{n}$ is the gyromagnetic ratio of the neutron and $\hbar$ the reduced Planck's constant.
The sign of the EDM effect $\Delta\omega_{\mathrm{EDM}}$ depends on the relative orientation of \Evec and \Bvec fields.
The EDM result is gained from a large number of experimental cycles that each represent one measurement of \omegan.
The signature of an existing \dn is a correlation of \omegan with the relative orientation of \Evec and \Bvec which is periodically reversed.

Current nEDM experiments use UCN which have a low kinetic energy and undergo total reflection under all angles of incidence from suitable wall materials due to the coherent neutron-nucleus interaction \cite{Golub1991}.
This allows the confinement of UCN in material storage vessels to perform spin precession experiments with hundreds of seconds free precession time.

The nEDM experiment that was installed at PSI's high intensity UCN source \cite{Anghel2009,Lauss2011, Lauss2012, Lauss2014} used a one chamber setup to measure \omegan for UCN exposed to a parallel and anti-parallel field configuration sequentially in time.
Instabilities of the magnetic field $\delta B$ led to increased statistical uncertainties in the measurement of \omegan and limited the experimental sensitivity to \dn until a mercury co-magnetometer was introduced to the search for an nEDM \cite{Green1998}.
The co-magnetometer is based on spin polarized \magHg atoms that are admitted to the same volume as the UCN to measure simultaneously the magnetic field experienced by the neutrons.
Using the measured Larmor frequency \omegaHg of the \magHg atoms, \omegan can be corrected for magnetic field changes that occur from one measurement cycle to the next.

While the statistical sensitivity to \dn can be improved by increasing the number of UCN with new high density UCN sources \cite{Kirch2010}, the magnetic field monitoring has to be improved at least at the same scale.
Here we demonstrate a factor of five increased magnetic sensitivity of a laser-based \magHg co-magnetometer in direct comparison with its predecessor which utilized discharge bulbs as light sources. In order to be certain that the main reason for improvement was the choice of light source and not the Hg polarization, we kept the Hg polarization apparatus the same, while changing the magnetometry light source. We then constructed a semi-empirical model that encodes potential imperfections in the light source spectrum due to non-magnetometer (vanishing nuclear spin) isotopes of Hg as well as exotic effects that reduce the expected light-atom interaction cross section into two parameters alpha and beta. For an ideal source, such as a UV laser, these two parameters are unity. For a non-ideal light source, such as a discharge lamp, we show that these two parameters are considerably different than unity and, if unaccounted for, could be misinterpreted as low Hg atomic polarization. This was indeed the case before this work and led to fruitless efforts to increase the apparent Hg atomic polarization in a variety of ways. We show conclusively that the Hg atomic polarization is actually high, as expected, and that the spectrum of the UV laser behaves much more like as ideal source which directly led to improved magnetometer sensitivity.

\section{\label{sec:MagReq}\magHg co-magnetometer requirements}

We define the Larmor precession frequency of \magHg atoms in analogy to \eqref{eqn:omegan}
\begin{equation}
\omegaHg=\gammaHg |\Bvec|.
\end{equation}
The gyromagnetic ratios of \magHg atoms $\gammaHg /2 \pi =  \SI{7.5901152(62)}{\hertz / \micro\tesla}$ \cite{Afach2014} and neutrons $\gamma_\mathrm{n}/2 \pi = \SI{29.1646933(69)}{\hertz / \micro\tesla}$ \cite{PDG2016} result in corresponding Larmor frequencies in the $|\Bvec| \approx \SI{1}{\micro\tesla}$ magnetic field used in our nEDM spectrometer.

Magnetic field drifts, from one measurement cycle to the next, shift $\omegan$ but a corrected frequency $\omegan^*$
\begin{equation}
\omegan^* = \omegan - \frac{\gamman}{\gammaHg}\omegaHg,
\label{eqn:CorrectedUCNFreq}
\end{equation}
can be obtained from the magnetic field measured with the \magHg co-magnetometer.
Under ideal conditions $\omegan^*$ is independent of magnetic field changes but keeps its sensitivity to \dn.
As the UCNs and the Hg atoms form two independent magnetometer systems (up to the common magnetic field) the statistical uncertainties of \omegan and \omegaHg propagate to $\omegan^*$
\begin{eqnarray}
\delta\omegan^* &=&\sqrt{ \delta\omegan^2 + \left( \frac{\gamman}{\gammaHg}\delta\omegaHg \right)^2} \\
&=& \delta\omegan \, C
\label{eqn:CorrectedUCNFreq2}
\end{eqnarray}
 with
 \begin{eqnarray}
 \quad C = \sqrt{1 + \left( \frac{\delta\omegaHg / \omegaHg}{\delta\omegan / \omegan} \right)^2}
\end{eqnarray}
Here $C$ represents the increase of statistical uncertainties due to the correction of \omegan using \omegaHg.
An acceptable increase of not more than 5\% means that the relative uncertainty on \omegaHg has to fulfill
\begin{equation}
\frac{\delta\omegaHg}{\omegaHg} = \frac{\delta B}{|\Bvec|} < \frac{1}{3.1} \frac{\delta\omegan}{\omegan}.
\end{equation}

For one measurement cycle, the smallest statistical uncertainty on \omegan that the nEDM experiment at PSI achieved in the 2015 and 2016 data runs is given by $\delta\omegan / \omegan = \SI{0.25}{ppm} $.
This leads to a maximum acceptable statistical uncertainty on \omegaHg of $\delta\omegaHg / \omegaHg < \SI{0.08}{ppm}$. In order to achieve this goal the magnetometer needs to provide an absolute magnetometric resolution of $\delta B < \SI{80}{\fT}$ in one experimental cycle.
The eight times improved \dn sensitivity aimed for in the next generation nEDM experiment at PSI, n2EDM, requires the \magHg co-magnetometer to provide a measurement uncertainty below \SI{10}{\fT}.

\section{\label{sec:ExpSetup}Experimental setup}

\begin{figure}
	\centering
		\includegraphics[width=50mm]{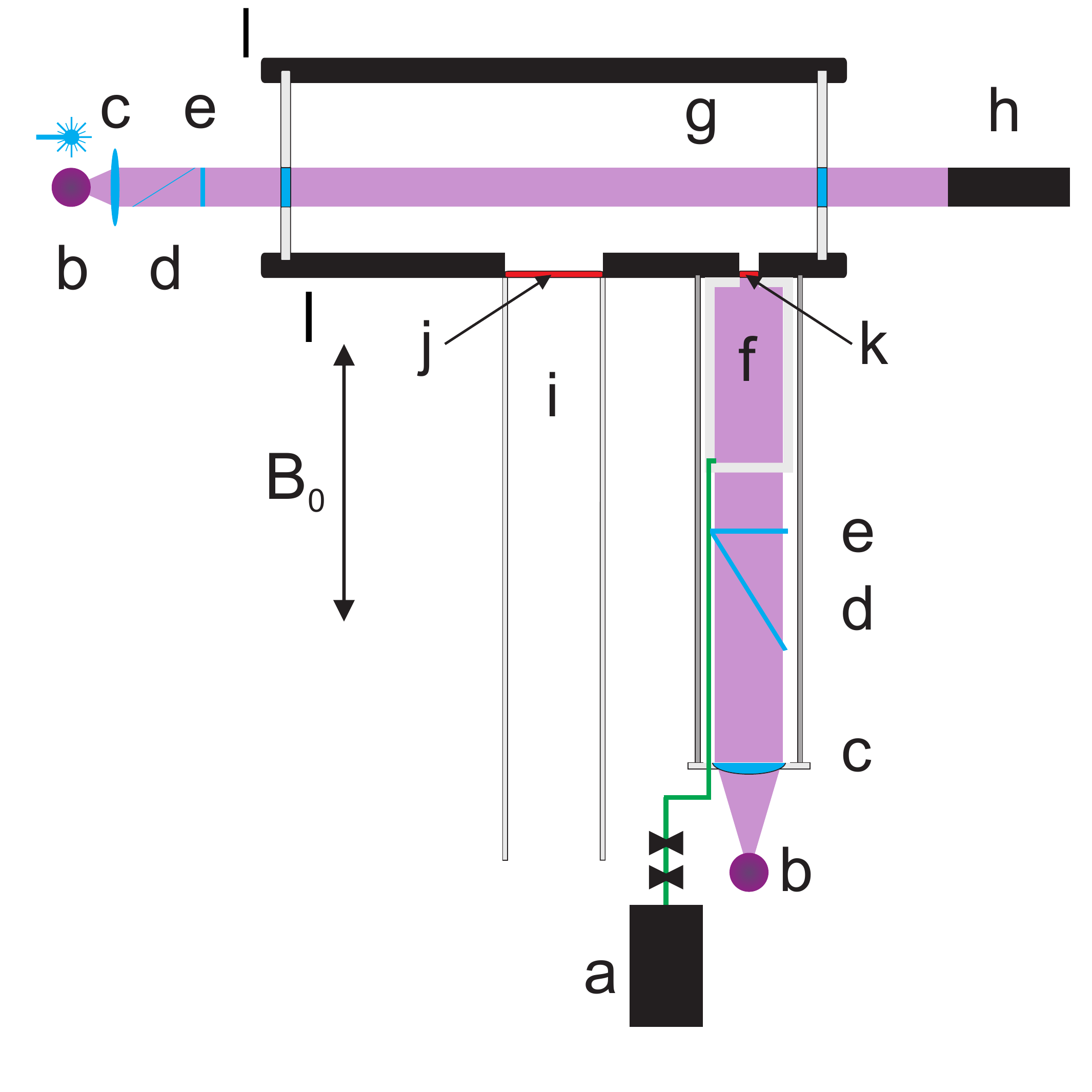}
	\caption[(Color online) The polarizing system of the \magHg co-magnetometer.]{The polarizing system of the $\magHg$ co-magnetometer in combination with a single beam absorptive detection scheme.  a) \magHg source, b) light source, c) lens, d) linear polarizer, e) $\lambda/4$-plate, f) polarisation chamber, g) UCN precession chamber, h) photo detector, i) UCN guide, j) UCN shutter, k) \magHg shutter, l) HV and ground electrodes. Not shown are the windows in the vacuum tank of the nEDM setup. The setup is to a large extent based on the hardware described in \cite{Green1998}.
	}
	\label{fig:PolarizationChamberAbsorption}
\end{figure}

Figure \ref{fig:PolarizationChamberAbsorption} shows the experimental setup and is restricted to the components relevant to the \magHg co-magnetometer of the nEDM experiment.
A detailed description of the complete nEDM setup at PSI can be found in \cite{Zenner2013}.
The magnetometer setup was mounted in the vacuum tank of the experiment and was exposed to a $\Bz\approx\SI{1}{\micro\tesla}$ bias magnetic field parallel to the axis of the cylindrical UCN storage chamber.
In the  following subsections we discuss the details of the magnetometer components relevant to its operation.
In subsection \ref{subsec:PropertiesMercury} we present the relevant properties of mercury. The properties of the different light sources in use are discussed in subsection \ref{subsec:LightSources}.
The $\magHg$ polarization chamber and the optical pumping process will be addressed in subsection \ref{subsec:ExpSetup:PolChamber}. Finally we discuss the UCN precession chamber where the magnetometry signal is generated and the light detection system in subsection \ref{subsec:ExpSetup:PreChamber}.

\subsection{\label{subsec:PropertiesMercury}Relevant Properties of Mercury}
The \magHg isotope has several distinct advantages over other atomic species that can possibly serve as co-magnetometer e.g.\ $^3\mathrm{He}$ or $^{129}\mathrm{Xe}$. The direct frequency shift of the \magHg Larmor frequency due to an EDM of the \magHg atoms is negligible as $\dHg \leq \SI{7.4E-30}{\ecm} \left(\SI{95}{\percent}\,\mathrm{CL}\right)$ is the most stringent limit on the EDM of any particle to date \cite{Graner2016}. The high vapor pressure at room temperature provides an easy path to produce an optically dense vapor column which does not affect the  UCN storage time due to the the low neutron absorption cross section of Hg. Furthermore, Hg is the only element for which direct optical detection of nuclear magnetic resonance (ODNMR) has been demonstrated without the use of laser cooling and trapping of atoms in their ground electronic state. Mercury is chemically very inert compared to the alkali metals, as e.g. used in the Cs-magnetometer array installed above and below the UCN precession chamber. Therefore Hg can be admitted to the UCN precession chamber as co-magnetometer species without the wall coatings required for alkali metals that would compromise the UCN storage properties of the UCN precession chamber.

\begin{figure}
	\centering
		\includegraphics[width=80mm]{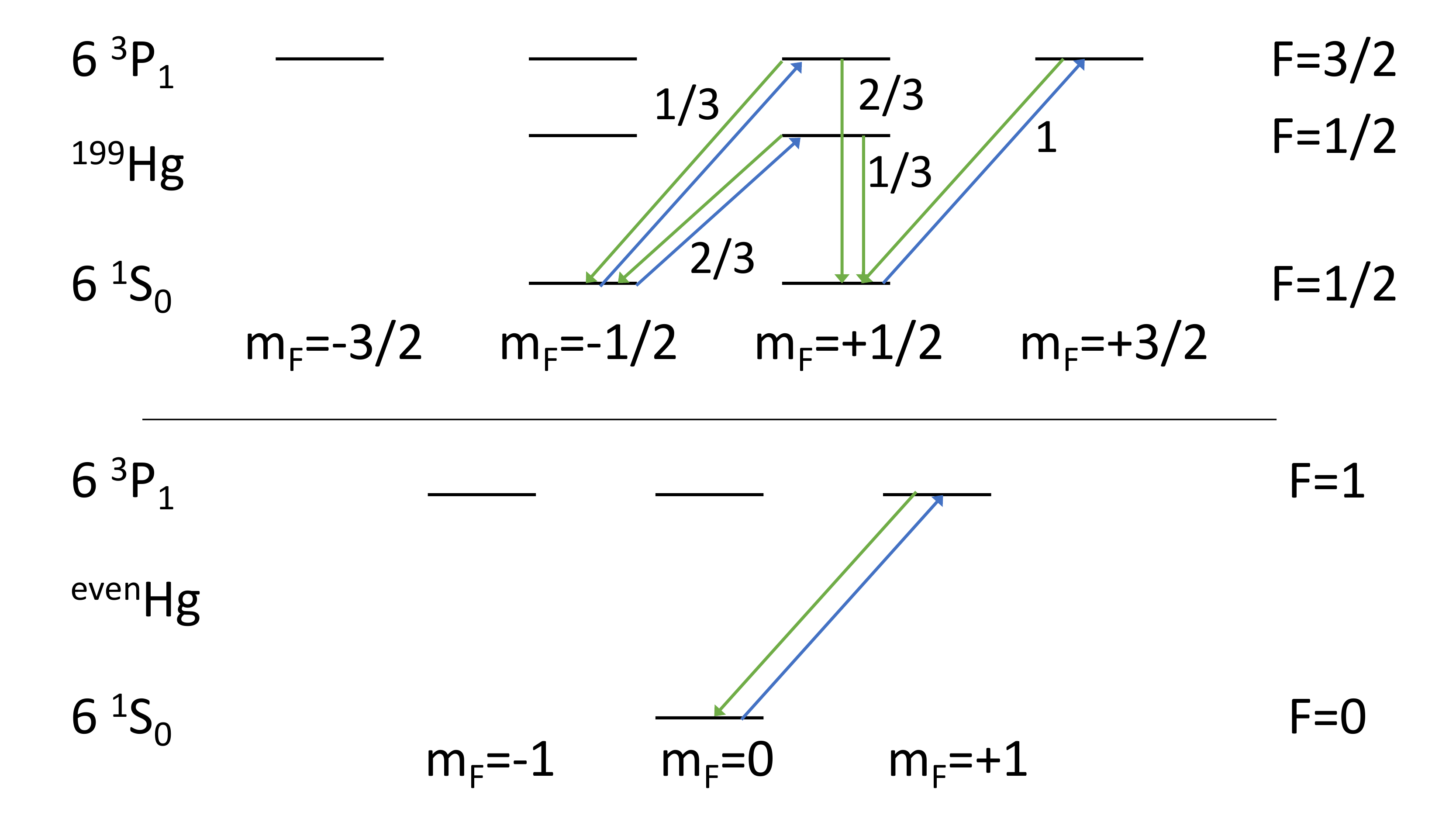}
	\caption[Lowest electronic states for the even and the $\magHg$ isotopes.]{Lowest electronic states for the $\magHg$ isotope (top) and even isotopes (bottom). The wavelength of the transition $6\,^1\mathrm{S}_0 \rightarrow 6\,^3\mathrm{P}_1$ is $\SI{253.7}{\nano\meter}$. The $\magHg$ isotope shows hyperfine splitting in the first excited state due to its nuclear spin $I=\frac{1}{2}$. The even isotopes possess no nuclear spin and thus no hyperfine structure. The squared Clebsch-Gordan-coefficient for each transition is given next to the transition arrow.}
	\label{fig:HgLevelStructure}
\end{figure}

The electronic ground state configuration of Hg is $[\mathrm{Xe}]4f^{14}5d^{10}6s^2$ in which the two $6s$ electrons are paired up in the  $6\,^1\mathrm{S}_0$ term to form a diamagnetic atom.
Figure \ref{fig:HgLevelStructure} shows the lowest lying electronic levels for the even isotopes and for \magHg.
The \SI{253.7}{\nano\meter} intercombination line connects the $6\,^1\mathrm{S}_0$ spin singlet ground state to the $6\,^3\mathrm{P}_1$ spin triplet excited state. The \magHg levels show hyperfine splitting due to the nuclear spin $I=\frac{1}{2}$ while the electronic levels of the even isotopes remain unsplit due to $I=0$. The squared Clebsch-Gordan-coefficients that determine the coupling between the different $m_\mathrm{F}$ states are also given in Fig.~\ref{fig:HgLevelStructure}.
For right-circularly polarized light ($\Delta m_\mathrm{F}=+1$) the $6\,^1\mathrm{S}_0,m_\mathrm{F}=+\frac{1}{2}$ state in \magHg is a dark state for the $6\,^1\mathrm{S}_0,F=\frac{1}{2}\rightarrow 6\,^3\mathrm{P}_1,F=\frac{1}{2}$ transition as the atom cannot absorb the angular momentum of the photon.
By the same argument the $6\,^1\mathrm{S}_0,m_\mathrm{F}=-\frac{1}{2}$ state in \magHg is a dark state for left circularly polarized light ($\Delta m_\mathrm{F}=-1$).
This dependence of the absorption cross section on the relative orientation of the photon and atom angular momenta is the basis for the optical pumping  and the detection of spin precession as described in the next subsections.
Further physical properties of the \magHg enriched sample used in this study are given in \cite{Fertl2013}.

\subsection{\label{subsec:LightSources}Light sources}

Traditionally \lampHg discharge lamps have been used in the \magHg co-magnetometer of the nEDM experiment to provide the \SI{253.7}{\nano\meter} deep ultra-violet light necessary for the optical pump and probe cycle.

In recent years frequency quadrupled diode laser systems have become commercially available.
This study compares the performance of the \magHg co-magnetometer for these two fundamentally different types of light sources used to detect the spin precession of polarized \magHg atoms.

\subsubsection{\label{subsubsec:LightSources:Bulbs}\lampHg discharge bulbs}
The \lampHg discharge bulbs, glass blown from a quartz glass tube, have a flat head (diameter $\approx \SI{10}{\milli\meter}$) and an extension tail.
The bulbs were first evacuated, then filled with argon buffer gas and a small amount of metallic mercury,  enriched to \SI{90}{\percent} in $\lampHg$, before being sealed.
For operation, an electrodeless RF discharge is ignited in the bulb which is mounted inside a temperature stabilized microwave cavity running at \SI{2.4}{\giga\hertz}.
Mercury atoms are excited in the plasma and relax to the electronic ground state by emission of $\SI{253.7}{\nano\meter}$ resonant light (see Fig.~\ref{fig:HgLevelStructure}).
The emission line of $\lampHg$ overlaps within the Doppler width ($\nu_\mathrm{D}\approx \SI{1}{\giga\hertz}$ at $T=\SI{298}{\kelvin}$) with the $6\,^1\mathrm{S}_0,F=\frac{1}{2}\rightarrow 6\,^3\mathrm{P}_1,F=\frac{1}{2}$ transition in \magHg  but does not show hyperfine splitting ($I = 0$). Therefore a $\lampHg$ bulb is a convenient light source to manipulate the nuclear (total) spin of $\magHg$ atoms.
Under typical operation conditions the $\lampHg$ discharge bulb emits $\approx \SI{20}{\micro\watt}$ of UV light distributed over the emission lines of all the Hg isotopes contained in the bulb. The discharge bulbs normally operate close to the photon shot noise limit. A detailed discussion of the emitted light spectrum is given in Sec.\,\ref{enum:DisadvantagesHgBulbs}.

\subsubsection{\label{subsubsec:Lightsources:Laser}UV laser system}
Alternatively, the resonant light can be provided by a high power, deep ultra-violet, frequency quadrupled diode laser system (FHG, \cite{Toptica2013}) that has been commissioned at PSI as an accurately controllable light source.
The FHG delivers up to \SI{25}{\milli\watt} of optical power around \SI{253.7}{\nano\meter} and can be frequency tuned, mode hop free, over a frequency range wider than the distance between the two resonance lines in \magHg ($\Delta f \geq \SI{22}{\giga\hertz}$).
The laser was frequency stabilized to a MHz line width using a sub-Doppler Dichroic Atomic Vapor Laser Lock (SD-DAVLL, \cite{Wasik2002,Petelski2003}).
As there was no appropriate setup area for the laser available next to the nEDM experiment, the UV light has been transported in \SI{50}{\meter} long, solarization resistant, multi-mode fibers (diameter \SI{600}{\micro\meter} or \SI{200}{\micro\meter}, $\mathrm{NA}=\num{0.22}$ \cite{Thorlabs2013}) from a laser laboratory to the nEDM setup. Light power levels have be varied between \SI{2}{\micro\watt} and \SI{15}{\micro\watt} but kept as low as possible to slow the solarization process of the fibers.

\subsection{\label{subsec:ExpSetup:PolChamber}Optical pumping}
The optical pumping of the \magHg atoms was performed in the polarization chamber, a cylindrical volume (ID \SI{73}{mm}, length \SI{287}{mm}) coated with perfluorinated paraffin to reduce wall-induced spin relaxation \cite{Chowduri2013}, below the grounded electrode of the UCN precession chamber (see Fig.~\ref{fig:PolarizationChamberAbsorption}).
The Hg atoms were continuously produced in the $\magHg$ source (a) outside the magnetic shield of the nEDM experiment.
The source thermo-dissociates HgO, with a highly enriched \magHg content, into Hg and O atoms.
The Hg atoms travel via a thin tube (ID \SI{4}{mm}) to the polarization chamber ((f), $V_\mathrm{pol} = \SI{1000}{\cubic\centi\meter}$ volume) where they are illuminated with resonant light from a $\lampHg$ discharge bulb (b) installed outside of the magnetic shield. Atomic densities up to \SI{7E15}{\per\cubic\centi\meter} in the polarization chamber have been used in this study in order to achieve the desired light absorption in the UCN precession chamber.
The light enters the vacuum tank through a focusing lens ((c), $f=\SI{+140}{\milli\meter}$) which also acts as the vacuum window. The light beam travels (anti) parallel to the main magnetic field and passes a linear polarizer (d) and a quarter wave plate (e) with its axis under \SI{45}{\degree} relative to the plane of the linear polarizer
The spins of the $\magHg$ atoms become spin polarized (anti-)parallel to the magnetic field direction by optical pumping (\cite{Cagnac1961}, see Figure \ref{fig:HgLevelStructure}). As can be seen from Fig.~\ref{fig:HgLevelStructure} it takes on average three photons to transfer a \magHg atom from the $m_\mathrm{F}=-\frac{1}{2}$ state to the $m_\mathrm{F}=+\frac{1}{2}$ via the $6\,^1\mathrm{S}_0,F=\frac{1}{2}\rightarrow 6\,^3\mathrm{P}_1,F=\frac{1}{2}$ transition.
The illumination time is determined by the overall cycle length of an nEDM measurement cycle and is between \SI{100}{\second} and \SI{350}{\second}.

\subsection{\label{subsec:ExpSetup:PreChamber}Precession Chamber}
The magnetometry signal is created in the UCN precession chamber (Fig.~\ref{fig:PolarizationChamberAbsorption} g, diameter $d=\SI{470}{\milli\meter}$, height $l=\SI{120}{\milli\meter}$ ) which is formed by a bottom and a top electrode and an insulating cylinder made from polystyrene.
 The insulator ring is coated on the inside with deuterated polystyrene (dPS) \cite{Bodek2008}.
 The aluminum electrodes are coated with diamond-like-carbon which guarantees high quality storage properties for the UCN \cite{Atchison2006}.
 The detection light beam for the \magHg magnetometer is either provided by a second $\lampHg$ discharge bulb (b) or the new laser system outside of the magnetic shield. The light passes a collimating lens (c), a linear polarizer (d) and a quarter wave plate (e) (with its axis oriented under \SI{45}{\degree} relative to the plane of linear polarization) before it enters and leaves the UCN precession chamber through two UV grade fused silica (UVFS) windows in the insulator ring.
 These UVFS windows are coated on the inner side with UV transparent deuterated polyethylene (dPE) to provide  good neutron storage properties \cite{Bodek2008}.
 The probe light beam is detected on the photocathode of a photomultiplier tube (PMT).
The polarized Hg vapor is released into the precession chamber after the UCN have been filled and the UCN shutter (j) has been closed.
After the free precession period of the nEDM measurement cycle the Hg atoms are evacuated when the UCN shutter  is opened again to guide the UCN to the UCN detector \cite{Afach2015EPJA, Ban2016}.

\section{\label{sec:ExpSetup:FSP}Free spin precession signal}
\begin{figure}
\centering
\includegraphics[width=80mm]{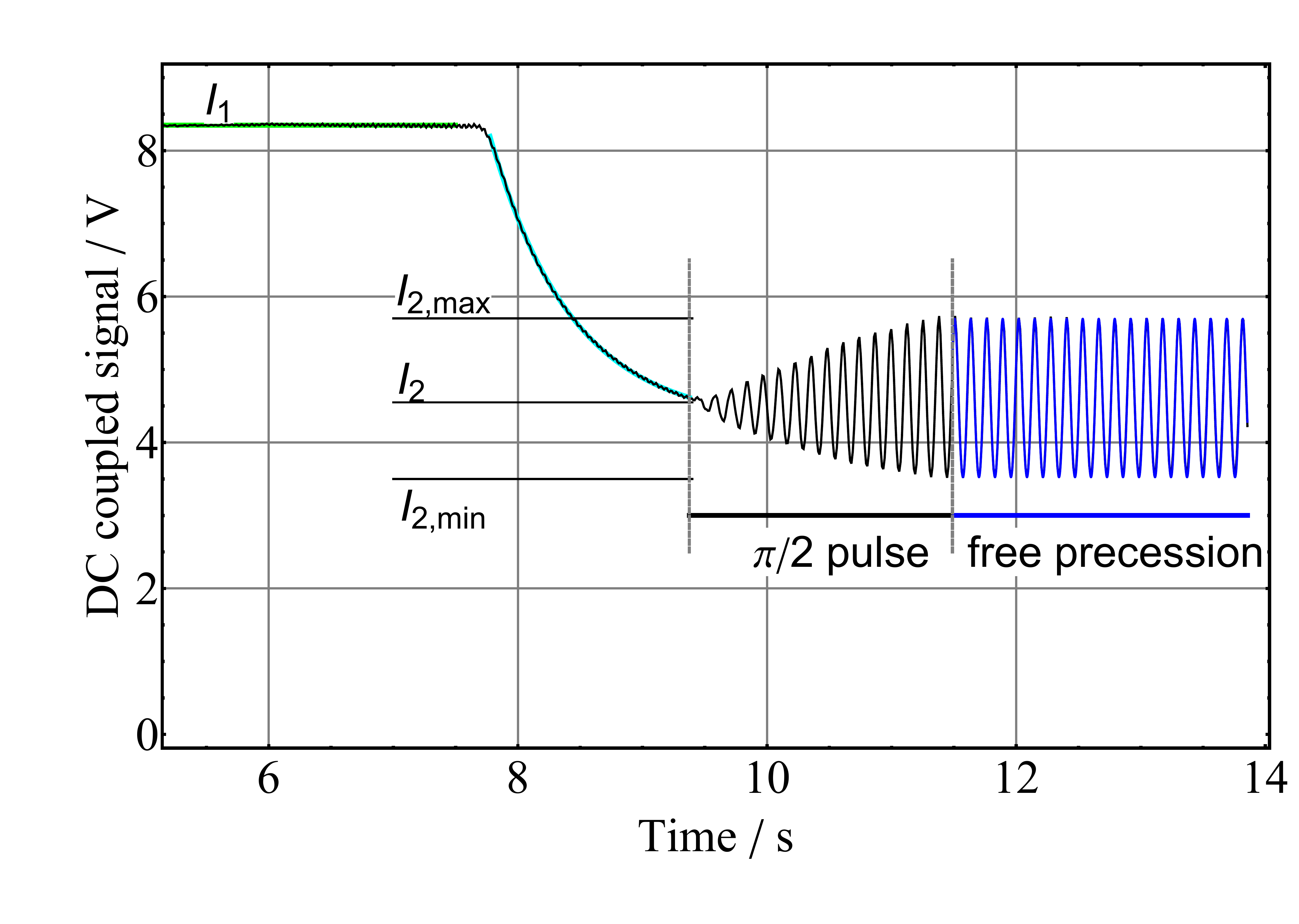}
\caption[DC-coupled Hg co-magnetometer signal]{A typical DC-coupled signal of the photon detector in the \magHg co-magnetometer. The $I_1$ signal level is determined without Hg in the precession chamber. After the release of Hg atoms into the chamber the $I_2$ signal level is measured. A $\pi/2$ pulse is applied to start the Larmor precession of the $\magHg$ atoms in the magnetic field. The observation of free induction decay starts at the end of the $\pi/2$ pulse .}
\label{fig:DCCoupledHgSignal}
\end{figure}

In order to measure the Larmor frequency of the \magHg atoms a free spin precession (FSP) signal is initialized by a $\pi/2$ pulse.
A typical signal of the \magHg co-magnetometer recorded with a DC coupled PMT is shown in Figure \ref{fig:DCCoupledHgSignal}.
The detected signal level $I$ is proportional to the photon flux impinging on the sensitive area of the PMT.
The signal level without Hg in the precession chamber, $I_1$, is determined shortly before the Hg-shutter valve is opened (at $t\approx\SI{8}{\second}$ in Fig.~\ref{fig:DCCoupledHgSignal}, (k) in Fig.~\ref{fig:PolarizationChamberAbsorption}) for \SI{2}{\second} to release polarized \magHg atoms into the precession chamber.
The detector signal is reduced to the level $I_2$ due to the light absorption by Hg atoms.
A circularly rotating magnetic field at \omegaHg is applied perpendicular to the main magnetic field for \SI{2}{\second} ($\pi/2$ pulse).
This magnetic resonance pulse flips the atomic spins into a superposition of their $m_\mathrm{F}=\pm1/2$ states which precesses in the plane perpendicular to the main magnetic field.
The observation time $T$ for the FSP starts at the end of the $\pi/2$ pulse.

A lower limit for the statistical uncertainties of the \magHg FSP magnetometer is given by the Cramer-Rao lower bound \cite{Gemmel2010} that reads for sufficiently high sampling rates:

\begin{equation}
\delta\mathrm{B} \geq \frac{\sqrt{12}}{\gammaHg\, \frac{a_s}{\rho} \, T^{3/2} } \underbrace{
\sqrt{\frac{e^{2/r}-1}{3 r^3 \left(\cosh \left(2/r\right)-1\right)-6 r}}
}_{D(r)}
\label{eqn:CramerRaoSensitivity}
\end{equation}
where $a_s = (I_\mathrm{2, max}-I_\mathrm{2, min})/2$ is the initial signal amplitude measured in V (see Fig.~\ref{fig:DCCoupledHgSignal} and \eqref{eqn:FertlSignal1}), $\rho$ the noise spectral density measured in $\mathrm{V}_\mathrm{RMS} / \mathrm{Hz}^{1/2}$, $T$ the observation time and $\gammaHg$ the gyromagnetic ratio of the $\magHg$ atoms.
In this calculation a white amplitude noise spectrum and a constant signal frequency is assumed.
The factor $D(r)$ takes into account the exponential decay of the signal with a characteristic decay time $\tau$ and depends only on the ratio $r=\tau/T$.
For typical values of our current nEDM experiment $T=\SI{180}{\second}$ and $\tau=\SI{100}{\second}$ it computes to $ D=\num{2.5}$.
Using this factor, the initial signal-to-noise-density-ratio (SNDR) $a_s/\rho$ for the \magHg co-magnetometer has to be larger than

\begin{equation}
\left(\frac{a_s}{\rho}\right)_\mathrm{nEDM} \geq  960\,\frac{\mathrm{V}}{\mathrm{V}/\sqrt{\mathrm{Hz}}}
\end{equation}
for the nEDM experiment and
\begin{equation}
\left(\frac{a_s}{\rho}\right)_\mathrm{n2EDM} \geq 7700\,\frac{\mathrm{V}}{\mathrm{V}/\sqrt{\mathrm{Hz}}}
\end{equation}
for the n2EDM experiment to achieve the sensitivity requirements outlined in Sec.~\ref{sec:MagReq}.
Since we use the noise density in the ratios above, they have a unit of $\sqrt{\mathrm{Hz}}$.
We chose this definition since it simplifies \eqref{eqn:CramerRaoSensitivity} and makes our results independent of measurement bandwidth.

\section{\label{sec:SignalModel}Signal model}

In routine operation, the performance of the \magHg co-magnetometer can be assessed and optimized by a small set of parameters like the measured light absorption in the Hg vapor, the FSP signal contrast, and the FSP signal decay time $\tau$.
An improvement of the Hg co-magnetometer can be achieved by increasing the atomic polarization $P$ of the Hg vapor.
To correctly determine the atomic polarization, a detailed model of the FSP signal generation is necessary.
A first FSP signal model for an absorption type measurement was introduced in \cite{Green1998}.
In our experiments it has become necessary to extend this initial model significantly since it is only valid for small FSP signal amplitudes and does not take into account the isotopic composition of the mercury vapor nor absorption cross section modifications as discussed in Subsection \ref{subsec:Disc:ContrastVsAbs}.
A typical FSP signal is shown in Fig.~\ref{fig:DCCoupledHgSignal}. We first derive the FSP signal amplitude and then give expressions for the atomic polarization as a function of the measured quantities.

\subsection{\label{subsec:SignalAmplitue}Signal amplitude}
The initial light level $I_1$ (see Fig.~\ref{fig:DCCoupledHgSignal}) is reduced according to Beer's law to
\begin{equation}
I_2 = \left(I_1 - I_{\mathrm{offset}}\right) e^{-n_\mathrm{199}\, \sigma_{\mathrm{199,1/2}} \,d \,\beta} + I_{\mathrm{offset}}
\label{eqn:FertlTransmittedBeforePulse}
\end{equation}
after the \magHg enriched vapor with a \magHg number density $n_\mathrm{199}$ was filled into the precession chamber.
The leakage of Hg atoms from the precession chamber was found to be negligible on the time scale of the measurement and thus $n_\mathrm{199}$ was assumed to be constant after the initial filling.
Since the detection light traverses the precession chamber along its diameter, $d$, perpendicular to the atomic spin polarization, the light absorption cross section for unpolarized monochromatic light has to be used.
For unpolarized light on resonance with the $F=1/2$ transition for \magHg the cross section is denoted by $\sigma_{\mathrm{199,1/2}}$ and its numerical value is $\sigma_{\mathrm{199,1/2}}=\SI{4.0E-13}{\square\centi\meter}$. The term $I_\mathrm{offset}$ accounts for light that cannot be absorbed by the Hg atoms in the enriched \magHg sample.
For example, this is light which is emitted by Hg isotopes in the \lampHg bulb different than the ones present in the enriched \magHg sample.
The laser light source tuned to the center of an absorption line closely approximates the ideal condition for which we define  $I_\mathrm{offset}=\SI{0}{\volt}$.
The correction factor $\beta$ is defined as
\begin{equation}
\beta = \frac{1}{n_\mathrm{199}\,\sigma_{\mathrm{199,1/2}}} \sum_{\mathrm{X}} n_\mathrm{X}\, \sigma_{\mathrm{X}}^*
\end{equation}
where $n_\mathrm{X}$ is the number density of Hg isotope $X$ (198, 199, ...) and $\sigma_{\mathrm{X}}^*$ is its the effective cross section which is the absorption cross section averaged over the probe light spectrum $\Phi(\nu)$
\begin{equation}
\sigma_{\mathrm{X}}^*=\frac{1}{\int_{-\infty}^\infty\Phi\left(\nu\right)d\nu}\int_{-\infty}^\infty \sigma_\mathrm{X}\left(\nu\right)\Phi\left(\nu\right)d\nu.
\end{equation}

Two different effects can contribute to the correction:
\begin{enumerate}
\item{Light absorption by isotopes other than $\magHg$ with modified cross section due to different or missing hyperfine splitting (e.g.\ $\sigma_{\mathrm{204}}= 3\sigma_{\mathrm{199,1/2}}$ or $\sigma_{\mathrm{199,3/2}}= 2\sigma_{\mathrm{199,1/2}}$).}
\item{Modifications of the effective light absorption cross-section of all Hg isotopes due to the light spectrum emitted by the \lampHg discharge bulb $\Phi\left(\nu\right)$.}
\label{enum:BetaFactorExplain}
\end{enumerate}
Two cases are of special interest for a monoisotopic  \magHg vapor.
For the narrow bandwidth laser light on resonance with the \magHg $F=1/2\,(F=3/2)$ line we find $\beta=1\, (2)$.
Assuming a Doppler-broadened absorption cross section
$\sigma_\mathrm{199,1/2}\left(\nu\right) \propto \exp \left(-\frac{(\nu-\nu_0)^2} {2\nu_{\mathrm{D}}^2}\right)$
and $\Phi\left(\nu\right) \propto \sigma_\mathrm{199,1/2}$ (as for an ideal discharge lamp without isotope shift) results in $\beta=1/\sqrt{2}$.
Here the center of the absorption line is given by $\nu_0$ and $\nu_{\mathrm{D}}$ is the Doppler width of the absorption line.

After the $\pi/2$ pulse (see Fig.~\ref{fig:DCCoupledHgSignal}) the atomic spins precess in the plane perpendicular to $B_0$ and thus alternate between being parallel and anti-parallel to the axis of the circular light polarization.
Therefore the absorption cross section for the mercury vapor with atomic polarization $P$ is modulated at the Larmor frequency, \omegaHg,
\begin{equation}
\sigma\left(t\right) = \left (\beta-\frac{P}{\alpha} \sin\left(\omegaHg t\right)\right)\sigma_{\mathrm{199,1/2}}.
\end{equation}
The factor $\alpha$ accounts for a possible reduction of the modulation amplitude due to a modification of the absorption cross section for the \magHg $F=1/2$ line
\begin{equation}
\alpha = \frac{\sigma_\mathrm{199,1/2}}{\sigma_\mathrm{199,1/2}^*}.
\end{equation}
For a monoisotopic \magHg vapor and the narrow band laser light source on resonance with the \magHg $F=1/2$ line $\alpha=1$, for an ideally Doppler-broadened light source (no isotope shift) $\alpha=\sqrt{2}$.

The time-dependent PMT signal is given by
\begin{equation}
I_2\left(t\right) = \left(I_1 - I_{\mathrm{offset}}\right) e^{-n_\mathrm{199}\, d \,\sigma\left(t\right)} + I_{\mathrm{offset}},
\label{eqn:FertlTransmittedAfterPulse}
\end{equation}
Independently of the optical thickness of the atomic vapor column ($n_\mathrm{199}\, d \,\sigma\left(t\right)$).
For an optically thin vapor column ($n_\mathrm{199}\, d \,\sigma\left(t\right) \ll 1$) the exponential function can be linearized while for a optically thick vapor column ($n_\mathrm{199}\, d \,\sigma\left(t\right) \gg 1$) the exponential term is the source of sigificant harmonic frequency generation. In order to extract the amplitude of the signal component oscillating at \omegaHg the exponential function in \eqref{eqn:FertlTransmittedAfterPulse} can be expanded in terms of $n_\mathrm{199}\, d \,\sigma\left(t\right)$ \cite{Fertl2013} in an infinite series.
While this expansion takes all the nonlinearity of the signal generation into account, it does not provide a closed expression for the FSP signal amplitude $a_s$.
To a very good approximation, the modulation signal amplitude $a_s$ can be approximated as half the difference between the maximum and the minimum PMT signal at the beginning of the FSP (see Fig.~\ref{fig:DCCoupledHgSignal})

\begin{equation}
\begin{aligned}
a_s = & \frac{I_\mathrm{2, max}-I_\mathrm{2, min}}{2} \\
      = & \frac{(I_1 - I_{\mathrm{offset}})}{2} \left(e^{-n_\mathrm{199} \, \sigma_{\mathrm{199,1/2}} \, d \left(\beta-\frac{P}{\alpha}\right)} \right. \\
      &\left. - e^{-n_\mathrm{199}\, \sigma_{\mathrm{199,1/2}} \, d \left(\beta+\frac{P}{\alpha}\right)} \right).
\end{aligned}
\label{eqn:FertlSignal1}
\end{equation}

\begin{figure}
	\centering
		\includegraphics[width=80mm]{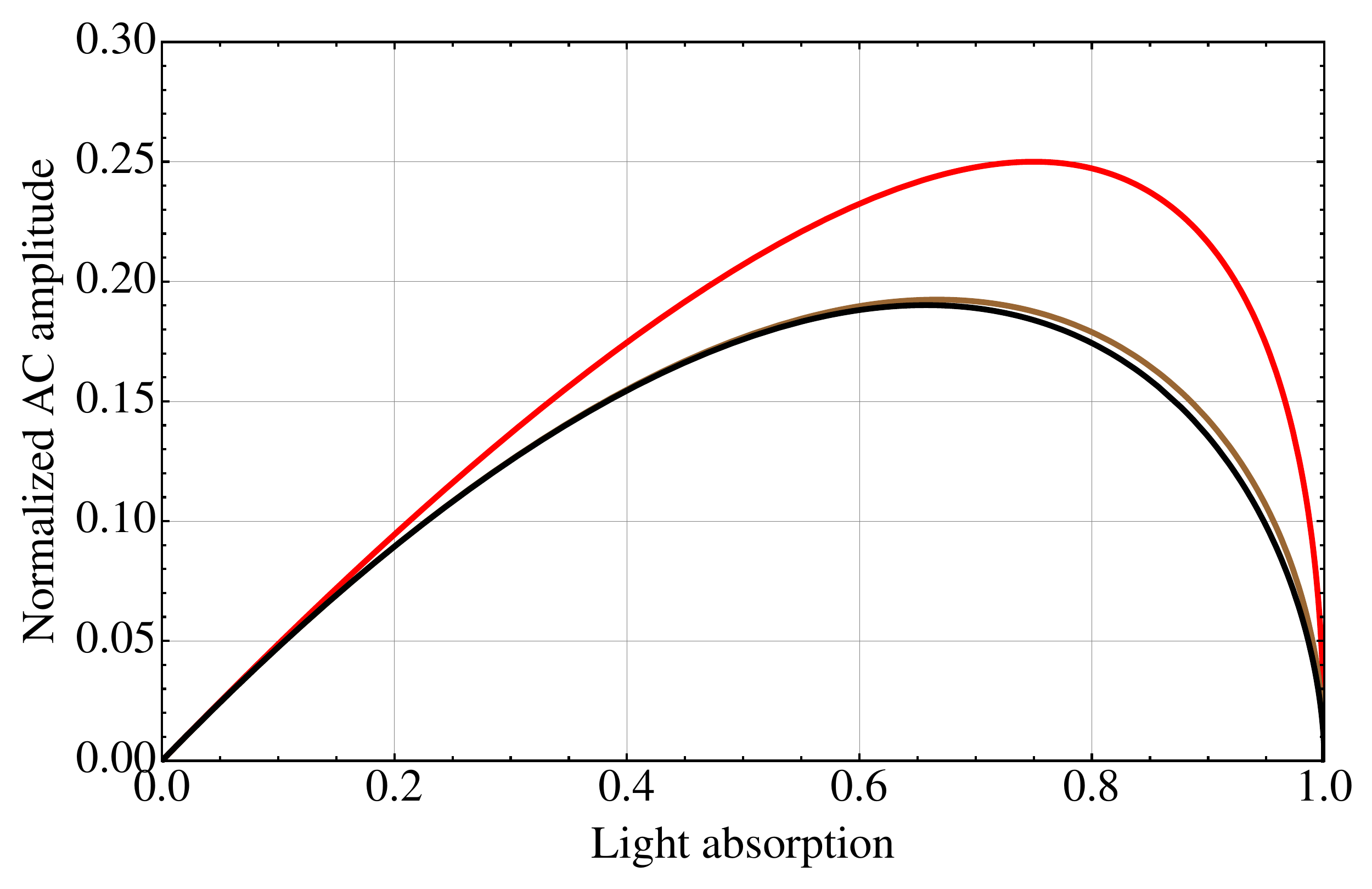}
	\caption[Model comparison]{A comparison of the signal amplitude predicted by three different models for fixed \magHg polarization $P=\SI{50}{\percent}$:
	The small signal model from \cite{Green1998} $\left(\PlotLineRed\right)$, the extended signal model \eqref{eqn:FertlSignalSinhMeasure}  $\left(\PlotLineBrown\right)$ and the exact signal model from \cite{Fertl2013} $\left(\PlotLineBlack\right)$. }
	\label{fig:ModelComaprison50Percent}
\end{figure}

We will refer to this approximation as the extended FSP signal model.
A comparison of the small amplitude model presented in \cite{Green1998}, the extended model of Eqn.~\eqref{eqn:FertlSignalSinhMeasure} and the exact model discussed in \cite{Fertl2013} is given in Fig.~\ref{fig:ModelComaprison50Percent} as a function of light absorption for a constant atomic polarization $P=\SI{50}{\percent}$.
The signal amplitude $a_s$ predicted by \eqref{eqn:FertlSignal1} deviates less than \SI{3}{\percent} from the exact model, except when $A^\mathrm{meas}\geq\SI{80}{\percent}$ (corresponding to an optical thickness $n_\mathrm{199}\, d \,\sigma\left(t\right) > \num{1.6} $.)
This discrepancy reflects that contributions to the amplitude modulation at \omegaHg caused by the terms proportional to $1/(2m+1)! \sin(2m+1)]\left(\omegaHg t\right)$ in the expansion of Eqn.~\eqref{eqn:FertlTransmittedAfterPulse} have been neglected. As the maximum
modulation amplitude occurs even with perfect atomic polarization below $A^\mathrm{meas}\geq\SI{80}{\percent}$ the extended model is always a sufficient signal approximation under the experimental conditions investigated in this study.

\subsection{\label{subsec:AtomicPolarization}Atomic polarization}
To extract the degree of atomic polarization $P$ of the \magHg atoms, $n_\mathrm{199}$ has to be determined from the measured light absorption value $A^{\mathrm{meas}}$
\begin{equation}
A^{\mathrm{meas}}= \frac{I_1-I_2}{I_1}.
\end{equation}
This is different from the expected light absorption if $I_\mathrm{offset}$ and light absorption by non \magHg isotopes has to be taken into account
\begin{equation}
A^{\mathrm{corr}} = \frac{I_1-I_2}{I_1-I_{\mathrm{offset}}} = 1- e^{-n_\mathrm{199}\,\sigma_{\mathrm{199,1/2}}\, d \, \beta}.
\label{eqn:FertlAbs}
\end{equation}

\noindent The light absorption due to the \magHg atoms alone is given by
\begin{equation}
A_{\magHg}=1-e^{-n_\mathrm{199}\,\sigma_{\mathrm{199,1/2}}\, d} = 1-\left(1-A^{\mathrm{corr}}\right)^{\frac{1}{\beta}}.
\end{equation}

\noindent The product $\beta \, n_\mathrm{199} \, \sigma_{\mathrm{199,1/2}} \,d$ can only be determined as a function of the a priori unknown amount of non-absorbable light:
\begin{equation}
\beta \, n_\mathrm{199}\, \sigma_{\mathrm{199,1/2}}\, d = - \ln\left(1-\frac{I_1-I_2}{I_1-I_{\mathrm{offset}}}\right).
\label{eqn:FertlAbsCoef}
\end{equation}

\noindent Substituting \eqref{eqn:FertlAbsCoef} in \eqref{eqn:FertlSignal1} gives for the signal amplitude $a_s$
\begin{equation}
a_s = \left(I_2 - I_{\mathrm{offset}}\right)\; \sinh\left(-\frac{P}{\alpha\beta}\ln\left(1-\frac{I_1-I_2}{I_1-I_{\mathrm{offset}}}\right)\right).
\label{eqn:FertlSignalSinhMeasure}
\end{equation}
\noindent Rearranging \eqref{eqn:FertlSignalSinhMeasure} for the atomic polarization $P$ gives
\begin{equation}
P=-\alpha\beta\frac{1}{\mathrm{ln}\left(1-\frac{I_1-I_2}{I_1-I_{\mathrm{offset}}}\right)}\arcsinh\left(\frac{a_s}{I_2-I_{\mathrm{offset}}}\right).
\label{eqn:FertlPol}
\end{equation}

From this expression it is clear that unaccounted changes of $\alpha$ or $\beta$ can be misinterpreted as changes of the atomic polarization. A particularly simple example is the intensity dependent light spectrum of the \lampHg lamps which will be discussed in Sec.\ref{enum:DisadvantagesHgBulbs}.

In the ideal case of $I_{\mathrm{offset}}=0$, $\alpha=1$ and $\beta=1$ the formula for the atomic polarization $P$ simplifies to
\begin{equation}
\begin{aligned}
P=&-\frac{1}{\ln\left(1-A^{\mathrm{meas}}\right)}\arcsinh\left(\frac{a_s}{I_2}\right)\\
=&-\frac{1}{\ln\left(1-A^{\mathrm{meas}}\right)}\arcsinh\left(\frac{a_s}{I_1\left(1-A^{\mathrm{meas}}\right)}\right).
\end{aligned}
\label{eqn:FertlPolSimple}
\end{equation}

Using $\arcsinh\left(x\right) = \ln\left(x+\sqrt{1+x^2}\right) $ in the limiting case of small signal contrast $x = a_s/(I_1(1-A^{\mathrm{meas}}))\ll1$ the expression for the atomic polarization as given in \cite{Green1998}
\begin{equation}
P=-\frac{\ln\left(1+\frac{a_s}{I_1\left(1-A^{\mathrm{meas}}\right)}\right)}{\ln\left(1-A^{\mathrm{meas}}\right)}
\label{eqn:RocciaPol}
\end{equation}
is recovered.

\section{\label{sec:Measurements}Measurement Configurations}

To compare the \magHg co-magnetometer signal recorded with the $\lampHg$ discharge bulb and the UV laser, three measurement runs were performed with an identical setup for optical pumping to polarize the \magHg atoms.
The number of Hg atoms, and thus the light absorption in the precession chamber, was controlled by varying the temperature of the HgO source which changes the dissociation rate of HgO. The three different detection light configurations were:
\begin{enumerate}

\item{\textbf{Configuration A:} UV laser light with the laser frequency stabilized to the $\magHg\,6\, ^1\mathrm{S}_0 \rightarrow 6\, ^3\mathrm{P}_1 \; F=1/2$ line ($F=1/2$ line).}
\item{\textbf{Configuration B:} UV laser light with the laser frequency stabilized to the $\magHg\,6\, ^1\mathrm{S}_0 \rightarrow 6\, ^3\mathrm{P}_1 \; F=3/2$ line ($F=3/2$ line).
}
\item{\textbf{Configuration C}: A $\lampHg$ discharge bulb temperature stabilized to \SI{40}{\degreeCelsius} as described in \cite{Roccia2009}. 
}
\label{enum:DifferentLightDetectionConfigurations}
\end{enumerate}
The transmitted light beam was detected with a solar-blind PMT (Hamamatsu 6834) outside of the nEDM vacuum tank.
The FSP signals were recorded with the DAQ system of the nEDM experiment.
The gain of the PMT was adjusted such that the DC signal values $I_1$ and $I_2$ could be digitized with a 12-bit resolution (10 V range) analog-to-digital converter (ADC).
The ac signal component was digitized after a second order multiple feedback bandpass filter (Allen and Key topology, measured center frequency $f_\mathrm{c}$=\SI{7.85}{\hertz} and quality factor $Q=\num{5.06}$) at a sampling rate of \SI{100}{\hertz}. A variable gain amplifier permits use of the full input range of a 16-bit ADC (\SI{3.3}{\volt} ADC range) for varying signal contrast. The variable gain factor was calibrated with a signal generator supplying a sine wave of \SI{7.85}{\hertz} and known amplitude to the PMT input of the nEDM DAQ system.

\section{\label{sec:DataAnalysis}Data Analysis}

The initial signal amplitude and the signal decay time of the FSP were determined from the time domain while the noise density was extracted from the frequency domain after Fourier transform (FT).
To suppress filter charging effects  due to the analog bandpass filter, the first second of the recorded FSP signal was discarded.
To extract the initial FSP signal amplitude, we perform a fit to the first second of the remaining digitized ac part of the FSP signal using the signal function
\begin{equation}
A_0 \cos \left( 2\pi \nu t +\phi_\mathrm{0}\right)+C_{\mathrm{ADC}},
\end{equation}
with a constant amplitude $A_0$, the \magHg precession frequency $\nu$, a signal phase $\phi_\mathrm{0}$, and $C_{\mathrm{ADC}}$ the middle ADC bin.
After the analog band pass filter, the amplitude of the residual contribution of the second harmonic $2 \omegaHg$ (caused by the non-linear nature of Beer's law) is typically on the order of $ 10^{-3}$ and can be neglected in the determination of the atomic polarization and the signal contrast. But the second harmonic has to be taken into account for the frequency determination as it can create a so-called \textsl{early-to-late} effect. The second harmonic decays with twice the decay rate of the fundamental signal. At early times the second harmonic contribution affects the frequency determined by a single frequency fit. In addition the analog band pass filter introduces a phase delay between the input and the output signal. This phase shift is $\approx \pi/2$ at $\omegaHg$ and $\pi$ at $2 \omegaHg$. Therefore the phase of the second harmonic has to be a free parameter in the frequency extraction fit.

\begin{figure}
	\centering
		\includegraphics[width=80mm]{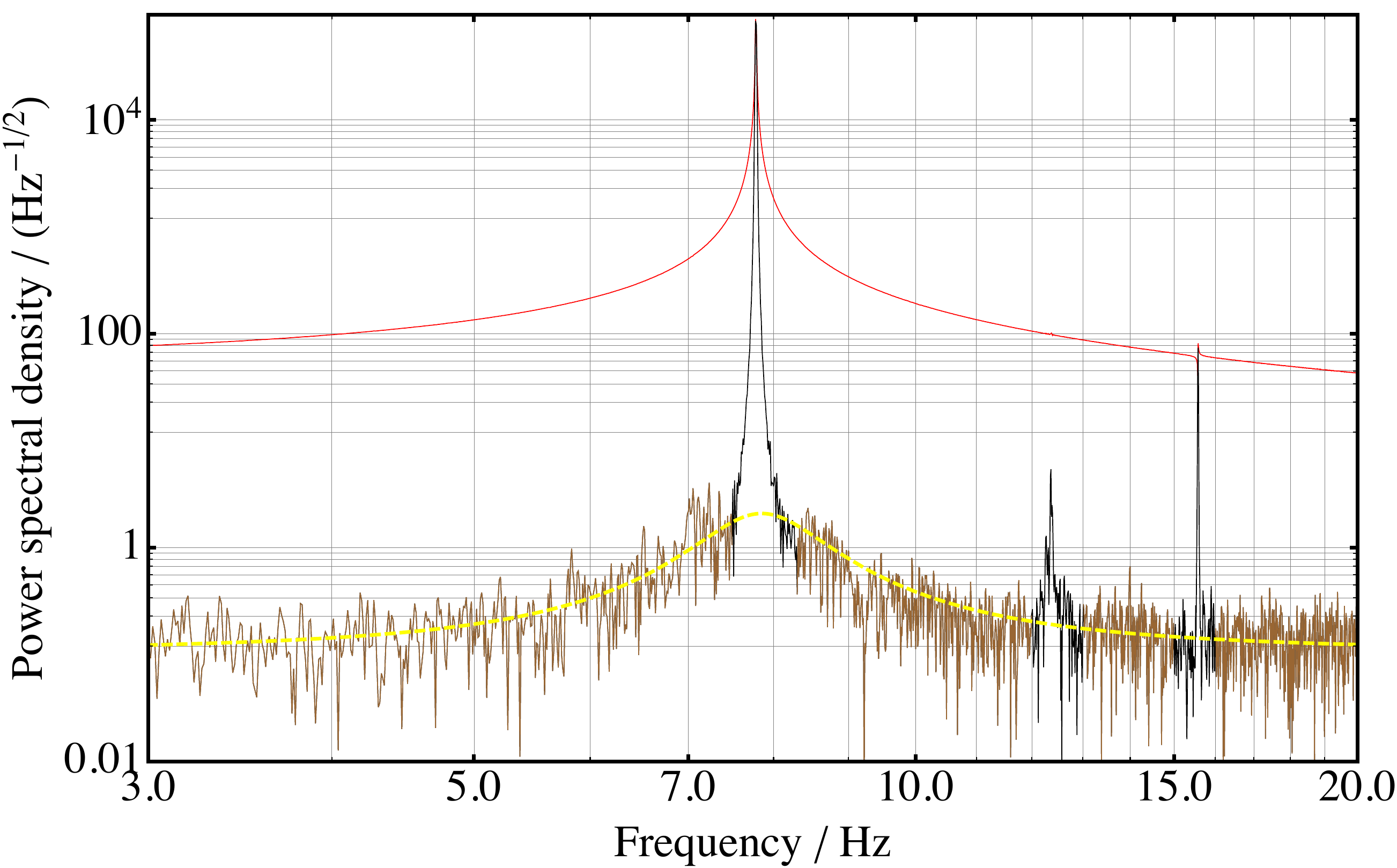}
	\caption[Power spectral density of nEDM Run 9494, cycle 123]{The power spectral density (PSD) of a typical FSP signal with and without signal windowing. While the PSD of the non-windowed FSP signal (\PlotLineRed) is completely dominated by the main peak around its center, one can extract the noise floor in the vicinity of the peak from the PSD of the windowed FSP signal (\PlotLineBlack). The transmission function of the analog bandpass filter is fitted to the windowed FSP signal (excluding regions of well peaked frequency components, (\PlotLineBrown)) to obtain an estimate of the PSD at the position of the main peak (\PlotLineYellowDashed) 
	.}
	\label{fig:FittedPSD_All_6494_123}
\end{figure}

To determine the SNDR at the signal frequency the power spectral density of the noise below the signal peak can be extracted via a discrete Fourier transform (DFT, see Fig.~\ref{fig:FittedPSD_All_6494_123}).
To extract the noise density under the signal peak a Hann (Hanning) window is applied to the FSP signal before the DFT.
The Hann window represents a good compromise between broadening the peaks and suppressing the leakage of spectral power into the frequency bins around the center peak which can be much larger than the noise floor in those bins.
The DFT with and without windowing are shown in Figure \ref{fig:FittedPSD_All_6494_123}.
The measured transmission function of the analogue band pass filter with a constant white noise background is fitted to the PSD of the windowed signal outside of well defined signal regions and used to determine the power spectral density of the noise at the signal frequency.

\section{\label{sec:Disc}Discussion}

\subsection{\label{subsec:Disc:ContrastVsAbs}Signal contrast}
\begin{figure}
  \centering
  \includegraphics[width=80mm]{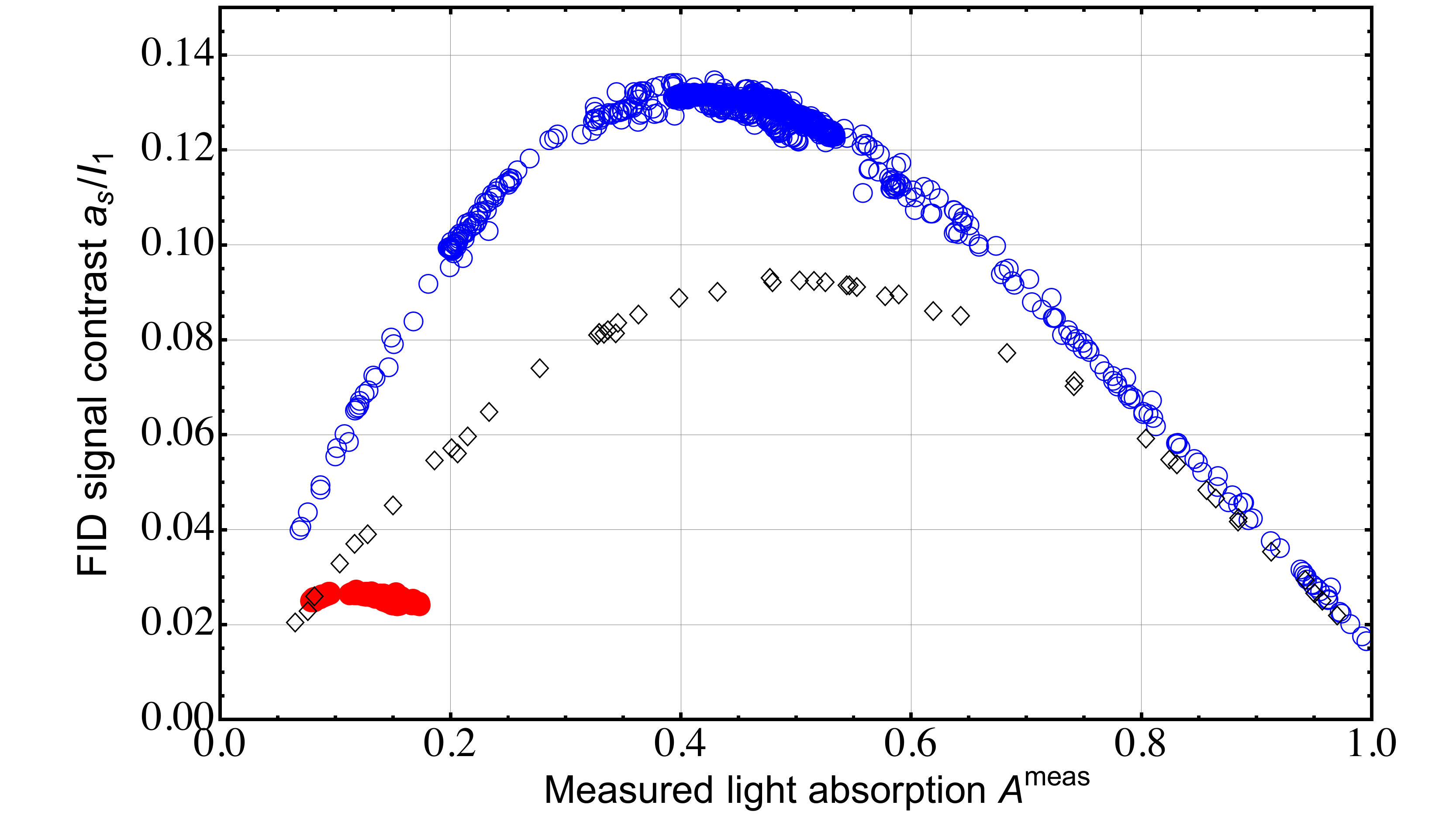}
  \caption[AC signal contrast for different readout light configurations.]{Normalized ac signal components $a_\mathrm{s}/I_1$  measured for three different readout light configurations as a function of he measured light absorption $A^\mathrm{meas}$: A) UV laser system frequency stabilized to the $\magHg\, F=1/2$ transition (\OpenCircleBlue) or B) the $\magHg\, F=3/2$ transition(\OpenDiamondBlack), C) $\lampHg$ discharge lamp (\FilledCircleRed) temperature stabilized to \SI{40}{\degreeCelsius}.}
  \label{fig:RawDataAmplitude}
\end{figure}

The signal contrast $a_\mathrm{s}/I_1$ is a convenient parameter to judge the \magHg co-magnetometer performance during regular operation since this ratio has to have a maximum between zero light absorption ($A^{\mathrm{meas}}=0$) and a completely opaque Hg vapour $A^{\mathrm{meas}}=1$.
For $A^{\mathrm{meas}}=0$ there are no atoms in the precession chamber that could contribute to the FSP signal.
On the other hand no photons reach the detector for $A^{\mathrm{meas}}=1$ and thus no FSP signal can be recorded.
The exact Hg density that yields the best signal contrast has to be determined experimentally.
This was done by slowly changing the temperature of the Hg source while recording a series of FSP signals.
Figure~\ref{fig:RawDataAmplitude} shows the signal contrast as a function of the measured light absorption $A^{\mathrm{meas}}$ as obtained for the three different measurement configurations.

For configuration A a value of $(a_\mathrm{s}/I_1)_{\mathrm{F=}1/2}^{\mathrm{laser}} = \SI{13.1}{\percent}$ at $A^{\mathrm{meas}}\approx\SI{41.5}{\percent}$ was found.
In configuration B $(a_\mathrm{s}/I_1)_\mathrm{F=3/2}^{\mathrm{laser}} = \SI{9.2}{\percent}$ was found at $A^{\mathrm{meas}}\approx\SI{49.2}{\percent}$.
The difference in contrast and in the position of the maximum between configurations A and B can be understood from the \magHg hyperfine splitting shown in Figure \ref{fig:HgLevelStructure}.
In the best case (\SI{100}{\percent} atomic polarization, \SI{100}{\percent} circular light polarization) the light absorption cross-section shows a modulation between 0 and $2 \sigma_{\mathrm{199,1/2}}$ for the $F=1/2$ line while the modulation is between $\sigma_{\mathrm{199,1/2}}$ and $3\sigma_{\mathrm{199,1/2}}$ for the $F=3/2$ line.
The polarization independent part of the light absorption cross-section does not contribute to the ac signal component but reduces the overall transmitted light level.
Therefore the maximum signal contrast is smaller and shifted to larger light absorption values for the $F=3/2$ line.
For configuration C a maximum value of $(a_\mathrm{s}/I_1)^\mathrm{bulb}_{\SI{40}{\degreeCelsius}} = \SI{2.6}{\percent}$ was found at $A^{\mathrm{meas}}\approx\SI{11.1}{\percent}$.
This is similar to the results reported in \cite{Green1998}.
The following reasons have been identified to explain the much smaller signal contrast of the \lampHg lamp based setup:
\begin{enumerate}
\label{enum:DisadvantagesHgBulbs}
\item{The emission lines of the \lampHg lamps are Doppler-broadened, reducing the average light absorption cross-sections.}
\item{Self-absorption (radiation trapping) causes a strong deviation of the spectral emission profile from a simple Doppler-broadened line shape (``line reversal"). Extensive Monte-Carlo simulations of this effect have been performed in the context of cylindrical Hg-Ar discharge light bulbs \cite{Bavea2003}. A strong temperature dependence is found by the authors as the mean free path of the light in the discharge volume is inversely proportional to the number density of mercury atoms, which itself depends strongly on the coldest spot temperature of the discharge bulb. The emission profile resembles a double-hump structure with a minimum at the atomic transition line center, thus decreasing the effective absorption cross section for Hg atoms outside of the discharge volume. These MC simulations have been performed with mercury of natural isotopic composition. We expect an even stronger influence on the emission spectrum of our discharge lamps due to the high degree of enrichment in \lampHg.}
\item{The self-absorption depends on the geometry of the bulb and the buffer gas pressure, two parameters that are not well under control during the production process. Different bulbs can emit significantly different light spectra at the same operation temperature and RF excitation power.}
\item{Emission lines of minority Hg isotopes are present in the bulb's light spectrum. Due to the lower optical density, light emitted by the minority isotopes suffers less self-absorption in the lamp head and is thus emitted at higher intensities than expected only from the mercury isotope ratio. This causes a large $I_\mathrm{offset}$ level.}
\item{To avoid line reversal the bulbs have to be run at sub-room-temperature. This lowers the density of the Hg atoms in the plasma and thus limits the achievable light output to several microwatt for the bulbs in use.}
\end{enumerate}

For the reasons given above the spectrum of the discharge lamps depends on many parameters.
In addition we found a dependence on the microwave power and the cavity mode that is most efficiently driven which is a function of the plasma impedance distribution.
We observed discrete jumps in the emission spectrum that are most likely due to changes in the preferred cavity mode.
In the setup of the neutron EDM experiment it was not possible to measure a spectrum.
Offline spectral analysis was not attempted since the conditions in the experiment could not be reliably reproduced.
Instead the lamp parameters were optimized to give the best signal for the Hg magnetometer.
For that reason we cannot give a more detailed description of the spectrum under the relevant conditions.

A detailed MC simulation of the not exactly known, but complicated bulb shape used in the nEDM experiment was not attempted.
But strong evidence that a combination of these factors explains the more than fivefold increase of 
signal contrast that was achieved with the UV laser frequency stabilized to the $\magHg\,6\, ^1\mathrm{S}_0 \rightarrow 6\, ^3\mathrm{P}_1 \; F=1/2$ line compared to the $\lampHg$ bulb
configuration is presented in the next section.

\subsection{\label{subsec:Disc:PolVsAbs}Polarization vs. Absorption}

In the past, several studies have been initiated to improve the signal contrast $a_\mathrm{s}/I_1$ of the \magHg co-magnetometer in the nEDM experiment in order to achieve the signal to noise ratio required for the next generation experiment, n2EDM. For example, the achieved signal contrast as reported in \cite{Green1998} was strongly dependent on the measured light absorption $A^\mathrm{meas}$ and much lower than achieved by the HgEDM experiment \cite{Swallows2013}. One working hypothesis was a high wall depolarization rate in the optical pumping chamber. This would prevent the Hg vapor from reaching a high equilibrium polarization. Several different wall coating materials were studied with the intention to achieve lower wall induced depolarization rates \cite{Horras2012}. Unfortunately no significant improvement was achieved. Another study aimed at increasing the useful photon flux of the \lampHg discharge bulb used for the optical pumping process. Again the achieved polarization depended on the specific \lampHg bulb used in the study and did not provide a conclusive explanation of the lower than expected signal contrast. All these studies relied on the simplified mathematical model given in \eqref{eqn:RocciaPol} to extract a value for the apparent atomic polarization $P$ from the detected signal contrast $a_\mathrm{s}/I_1$ and the measured light absorption $A^\mathrm{meas}$.

\begin{figure}
 	\centering
		\includegraphics[width=80mm]{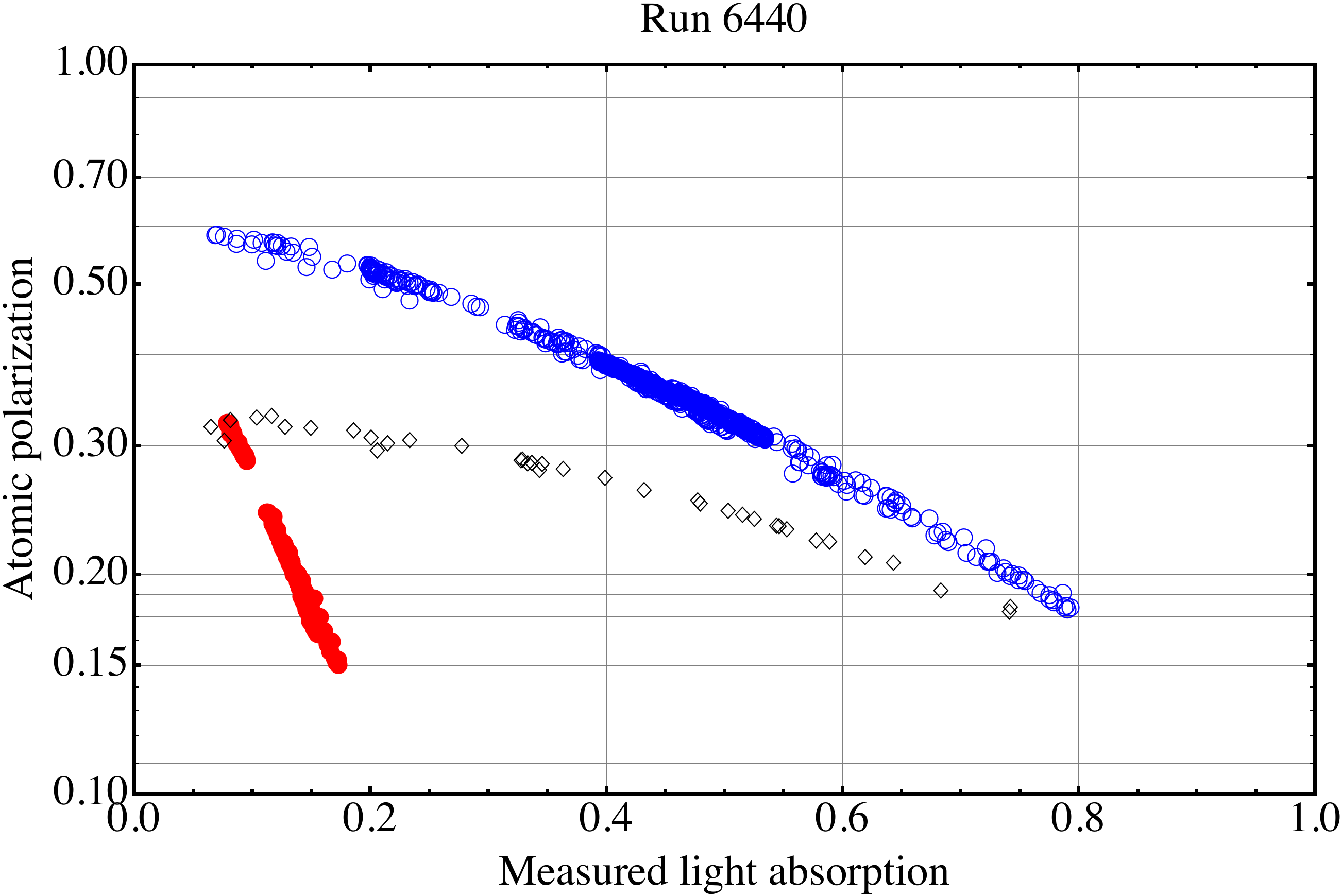}
	 	\caption[Atomic polarization according to the small amplitude model.]{Atomic polarization for the data sets given in Figure \ref{fig:RawDataAmplitude} determined with the small signal amplitude model, as given by \eqref{eqn:RocciaPol} \cite{Green1998}, for $A^\mathrm{meas}\leq\SI{80}{\percent}$. The three different readout light configurations are: UV laser frequency stabilized to the $\magHg\, F=1/2$ transition (\OpenCircleBlue, \SI{212}{\second} optical pumping time (opt)) or the $\magHg\, F=3/2$ transition (\OpenDiamondBlack, \SI{212}{\second} opt) and the $\lampHg$ discharge lamp (\FilledCircleRed, \SI{140}{\second} opt) temperature stabilized to \SI{40}{\degreeCelsius}.}
 	\label{fig:PolCompRALLogCut90Percent_6440_6444_6453_6454_6461_6462}
\end{figure}

\begin{figure}
	\centering
	\includegraphics[width=80mm]{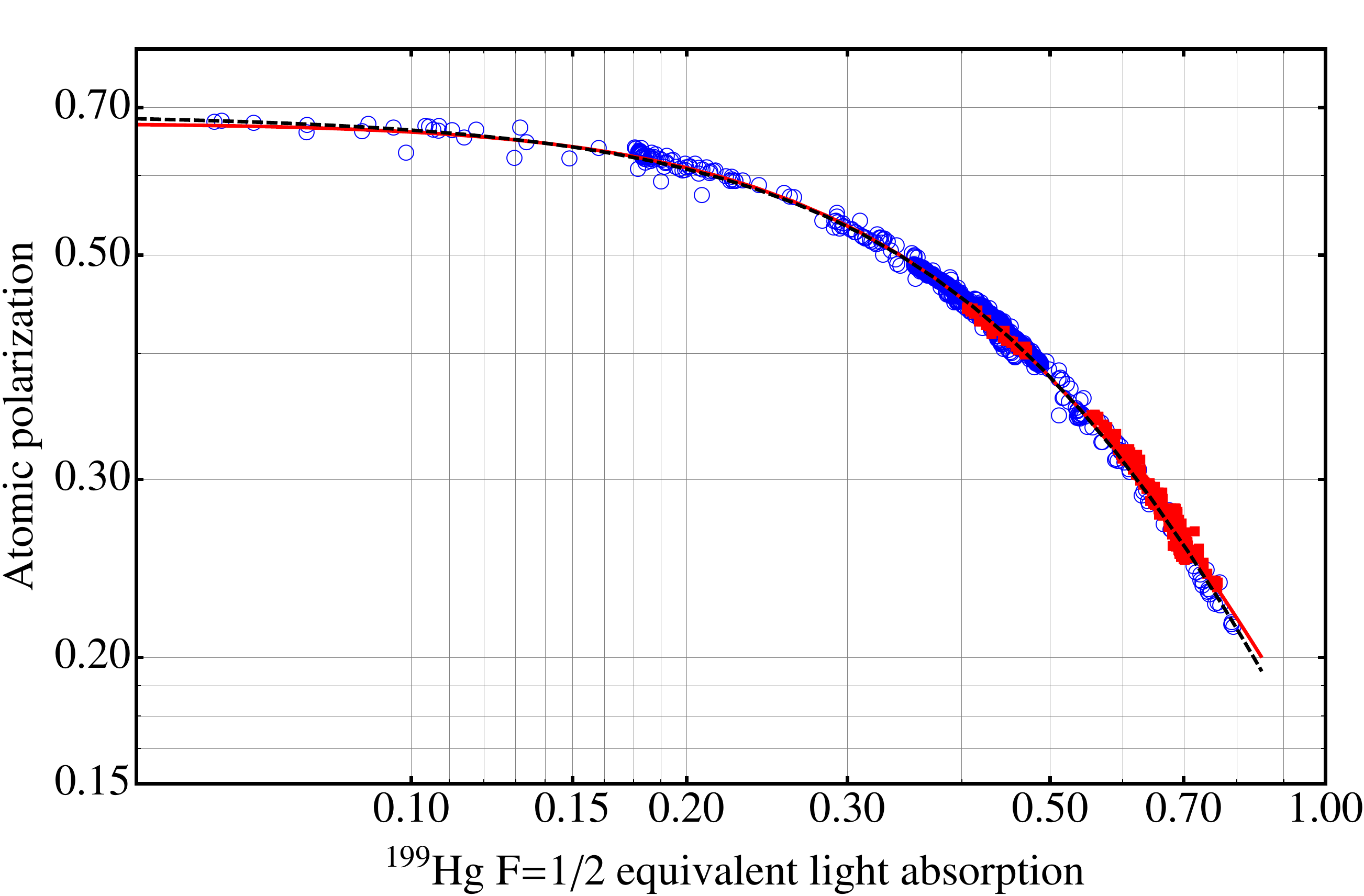}
 	\caption[Corrected atomic polarization vs light absorption.]{The parameters $I_\mathrm{offset}$, $\alpha$ and $\beta$ of the extended model can be adjusted in the analysis of the \lampHg discharge bulb based measurements (\FilledSquareRed) such that the determined atomic polarization shows the same scaling behavior as for the measurements performed with the laser frequency stabilized to the \magHg $F=1/2$ transition (\OpenCircleBlue). The optimum parameters to minimize the mean squared distance of the corrected lamp detected data set to the three parameter interpolation function (\PlotLineRed) are given in table \ref{tab:BulbCorrectionParameters}. The alternative four parameter interpolation function is also shown (\PlotLineBlackDashed)}
	\label{fig:20131025_PolarizationExtraction_LampVsLaserScalingBehavior}
\end{figure}

The data presented in subsection \ref{subsec:Disc:ContrastVsAbs} was recorded with identical operation conditions of the \lampHg bulb used for the initial optical pumping of the \magHg vapor.
The \magHg vapor was always polarized to its equilibrium value as was verified by changing the optical pumping time for configuration A between \SIrange{140}{212}{\second} without finding any difference in signal contrast.
This means that the preparation of the Hg atoms in all configurations (A, B, and C) was identical.
As a consequence one expects to observe the same relation between light absorption and atomic polarization no matter which light source is used to probe the spins.
Light absorption $A$ and atomic polarization $P$ should only depend on the processes in the polarization chamber that lead to the equilibrium spin polarization.
Since the pump light power was kept constant only the density of Hg atoms in the polarization chamber (controlled via the Hg source temperature) should influence the equilibrium.
This density is experimentally accessible as the light absorption $A$ which is thus chosen as the relevant x-axis in figures \ref{fig:PolCompRALLogCut90Percent_6440_6444_6453_6454_6461_6462} and \ref{fig:20131025_PolarizationExtraction_LampVsLaserScalingBehavior}.
The measurement result as presented in Fig.\,\ref{fig:PolCompRALLogCut90Percent_6440_6444_6453_6454_6461_6462} where the small signal amplitude model \eqref{eqn:RocciaPol} was applied to the data sets shown in Fig.~\ref{fig:RawDataAmplitude} shows a clear contradiction to this expectation.
The relation of $A^\mathrm{meas}$ and $P$ is clearly different for the three configurations due to the inaccurate extraction of those parameters in the small signal amplitude model.
Only data points for $A^\mathrm{meas}\leq\SI{80}{\percent}$ have been taken into account (see Sec.\ref{sec:SignalModel}).

These discrepancies have motivated the development of the extended signal model in order to understand the deviations between expectation and observation, and among the observations themselves.
If the extended model is plausible it must be possible to find the expected common relation of $A^\mathrm{meas}$ and $P$.
In order to verify this we chose configuration A (narrow linewidth laser light, locked to the \magHg $F=1/2$ transition) as the reference configuration.
Here the small linewidth of the laser justifies the assumption of $\alpha=1$ and parameter $\beta=1.14$ was calculated using the Hg vapor isotopic composition as given in the datasheet of the supplier of the enriched Hg sample.
The parameters in the extended model were then adjusted to find the same relation in configuration C.
 The obtained values for $P$ are shown in Fig.\,\ref{fig:20131025_PolarizationExtraction_LampVsLaserScalingBehavior} as a function of $A_{\magHg}$.
%
It was found that an empirical three parameter function
\begin{equation}
 P(A_{\magHg}) = \frac{1}{M_\mathrm{1}+M_\mathrm{2}\,A_{\magHg}+M_\mathrm{3}\,A_{\magHg}^2},
 \label{eqn:InterpolationFunctionForLampComparison}
\end{equation}
with $M_\mathrm{1}=1.49$, $M_\mathrm{2}=-0.266$ and $M_\mathrm{3}=5.16$, describes the scaling behavior of P vs. $A^{199}$ very well.
These parameters encode the details of the optical pumping process and the pre-polarization chamber surface properties which where not the target of this study.
In order to determine $\alpha$, $\beta$ and $I_\mathrm{offset}$ for configuration C all three parameters were varied such as to minimize the mean squared distances between the extracted atomic polarization and the heuristic interpolation function \eqref{eqn:InterpolationFunctionForLampComparison}.
The best agreement is obtained for  $\alpha=3.43(5)$, $\beta = 0.375(25)$, and $I_\mathrm{offset}=4.63(7)V$.
We estimate the uncertainty of $\alpha$, $\beta$, and $I_\mathrm{offset}$ as the deviation from the optimal parameter set as obtained from a four parameter interpolation function
\begin{equation}
\begin{split}
&P(A_{\magHg})= \\ &\frac{1}{M_\mathrm{1}+M_\mathrm{2}\,A_{\magHg}+M_\mathrm{3}\,A_{\magHg}^2+M_\mathrm{4}\,A_{\magHg}^3},
\end{split}
\end{equation}
with $M_\mathrm{1}=1.45$, $M_\mathrm{2}=0.191$, $M_\mathrm{3}=3.68$ and $M_\mathrm{4}=1.42$.
Both interpolation functions are shown in Fig.~\ref{fig:20131025_PolarizationExtraction_LampVsLaserScalingBehavior}.

The two data sets now show an excellent agreement of their scaling behavior.
Table \ref{tab:BulbCorrectionParameters} compares the extracted correction factors $\alpha$, $\beta$ and $I_\mathrm{offset}$ for the bulb light source and the laser. 
\begin{table}
\begin{center}
\begin{tabular}{|c |c | c | c|}\hline
parameter & $I_\mathrm{offset}/\mathrm{V}$ & $\alpha$  & $\beta$\\ \hline
laser & $0$ & $1$ & $1.14$\\\hline
lamp & $4.63(7)$ & $3.43(5)$ & $0.375(25)$\\\hline
\end{tabular}
\end{center}
\caption{Overview of the parameters $I_\mathrm{offset}$, $\alpha$, and $\beta$ used to the achieve the common atomic polarization vs.\ \magHg equivalent light absorption scaling behavior for measurements taken with the UV laser locked to the \magHg $F=1/2$ transition and with the Hg discharge lamp. $I_\mathrm{offset}$ parameterizes light that does not interact with the atoms in the precession chamber which is only the case for the lamp. $\alpha$ is a reduction factor for the efficiency with which the AC part of the signal due to the precessing \magHg atoms is recorded. $\beta$ describes the efficiency with which the DC part of the signal is recorded. In the case of the laser $\beta$ is larger than one due to isotopes present in the precession chamber that absorb stronger than \magHg while for the lamp $\beta$ is smaller than one because the broad spectrum more than compensates the increase due to isotopic composition.}
\label{tab:BulbCorrectionParameters}
\end{table}
The factor $\alpha=3.43(5)$ indicates a strong line reversal effect that decreases the effective absorption cross section drastically.
Caused by the high optical density for light emitted by \lampHg atoms in the bulb head only photons in the wings on the emission line escape efficiently.
But these photons have a drastically reduced absorption cross section compared to light on resonance with \magHg atoms.
The factor $\beta=0.375(25)$ indicates strong light absorption by isotopes other than \magHg.
These isotopes are also present in the Hg discharge bulb but with significantly reduced optical density for their resonance light.
Therefore light emitted by these isotopes does not suffer from a strong line reversal effect and can be absorbed by the atoms of the same residual isotopes present in the enriched \magHg vapor.
The atoms contribute to the number of photons but don't contribute to the ac signal component and thus reduce the signal to noise ratio.
The large value of $I_\mathrm{offset}$ indicates that nearly \SI{50}{\percent} of the dc ADC input range in configuration C was occupied by light that is not absorbed at all.
This light fraction does not contribute to the signal, but constitutes an extra source of noise.

We conclude that the extended signal model for the \magHg co-magnetometer signal provides a coherent analysis framework to take the isotopic composition of the enriched \magHg vapor and the readout light spectrum into account.
Furthermore, we have to conclude that the atomic polarization achieved in earlier experiments was already much higher than reported and efforts for improvements of the \magHg co-magnetometer might have been misguided by the unaccounted effects of the \lampHg discharge lamp spectrum.

\subsection{\label{subsec:Disc:SNR}Signal to Noise Ratio}

\begin{figure}
	\centering
		\includegraphics[width=80mm]{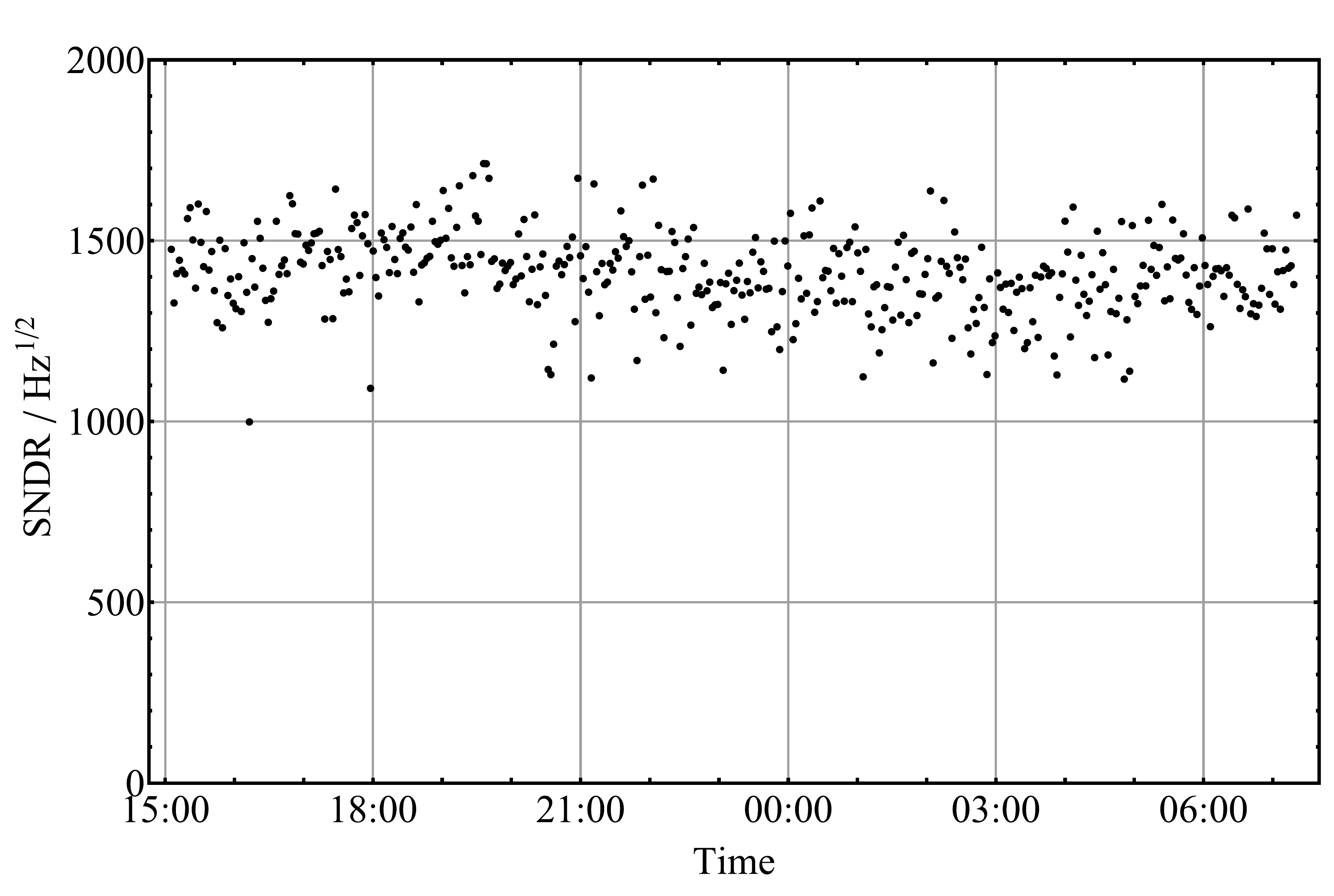}
		\caption[Achieved SNDR for \lampHg bulb detection light]
{The achieved signal-to-noise-density ratio for the \magHg magnetometer with a \lampHg discharge bulb at \SI{40}{\degreeCelsius} over a period of \SI{17}{\hour}.}
	\label{fig:20131108_1248_SNDR_extractionSNDR_6440_Time}
\end{figure}

\begin{figure}
	\centering
		\includegraphics[width=80mm]{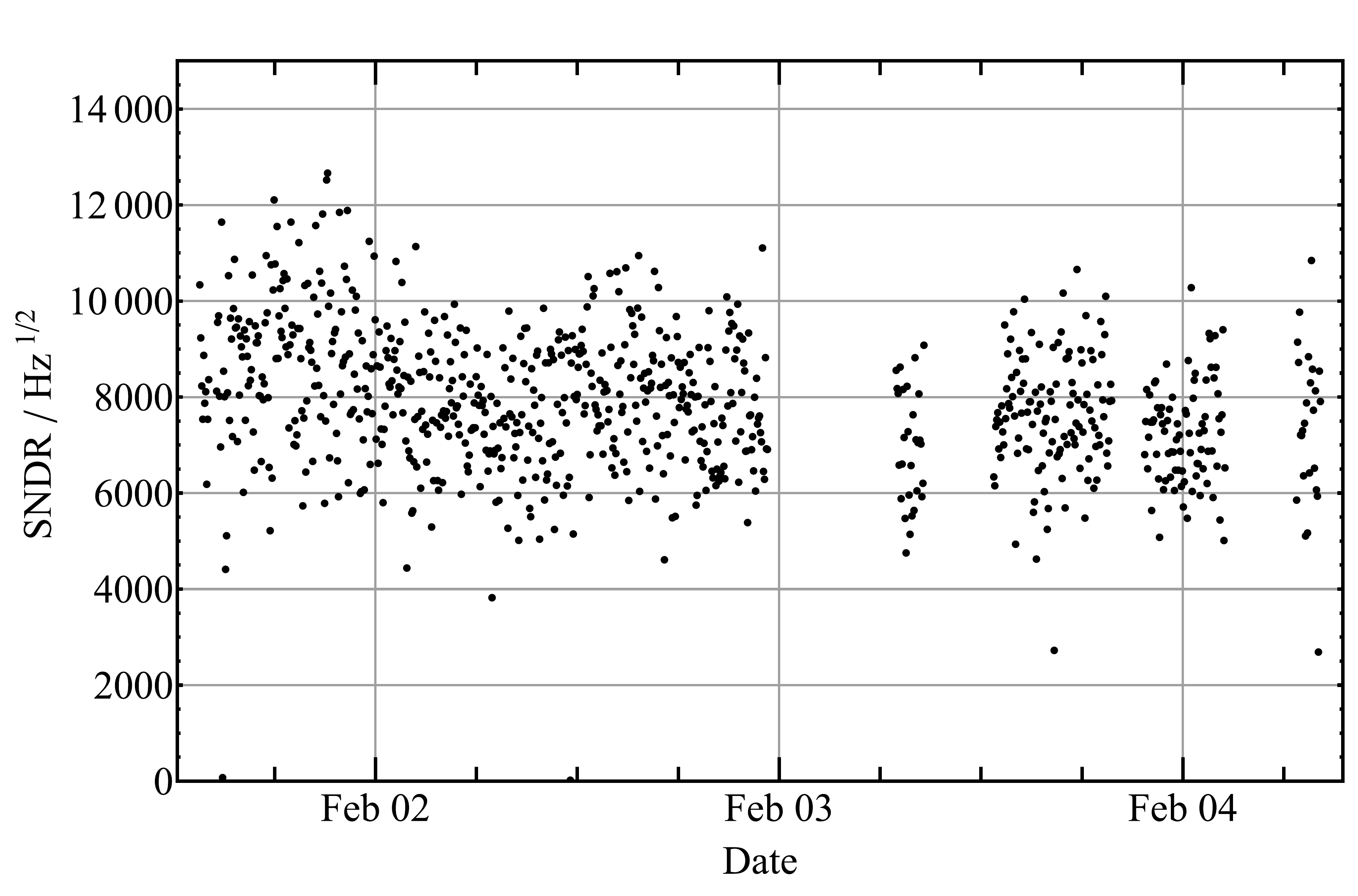}
		\caption[ Achieved SNDR for UV laser detection light]
{The achieved signal-to-noise-density ratio for the \magHg magnetometer with the UV laser frequency stabilized to the \magHg $F=1/2$ line over a period of more than \SI{66}{\hour}.
The frequency stabilization failed several times due to a strong weather change.
The change in ambient air pressure (by about \SI{30}{\milli\bar}) would cause an uncompensated shift of the laser frequency by about \SI{600}{\mega\hertz}.
The PID controller for the frequency stabilization reached its pre-set (software) limits of its control output. The PID controller can easily compensate such large changes with increased output limits.}
	\label{fig:20131108_1248_SNDR_extractionSNDR_6494_Time}
\end{figure}

The initial SNDR of each FSP has been determined for the typical running conditions in configurations A and C.
The SNDR achieved for the \lampHg bulb is $\approx \num{1410\pm110} \, \sqrt{\mathrm{Hz}}$.
A data taking sequence over the course of 17 hours is shown in Fig.~\ref{fig:20131108_1248_SNDR_extractionSNDR_6440_Time}.
For the UV laser system resonantly driving the \magHg $F=1/2$ transition we find initial $\mathrm{SNDR} = \num{7860\pm1380} \,\sqrt{\mathrm{Hz}}$.
We were able to operate the laser based system for more than one day without interruption as can be seen in Fig.~\ref{fig:20131108_1248_SNDR_extractionSNDR_6494_Time}.
During time periods where no measurement points are shown the laser dropped out of its wavelength-lock due to a (conservatively set) software limitation of the PID loop.

\section{\label{sec:Summary}Summary}

This work demonstrates the significant improvement in signal quality that can be achieved by using a laser system instead of a discharge lamp as the light source for \magHg co-magnetometer readout in experiments searching for a nEDM.
We could demonstrate that the lamp-based readout was sufficient for the PSI nEDM experiment in the 2015 and 2016 data runs.
The laser based readout offers an improvement in signal to noise ratio by a factor of 5.5.
An extended model of the FSP signal dependence on \magHg density was developed in order to understand the factors that contribute to this significant improvement.
The main conclusion is that the absorption cross section for the light generated by the discharge lamp is significantly smaller than for light resonant with the \magHg atoms.
In addition, the lamp generates a large amount of off-resonant UV light that does not contribute to the absorption signal but can induce, for example, intensity dependent Larmor frequency shifts (vector light shift).

The laser based readout was investigated as an option for the next generation nEDM experiment at PSI.
Assuming an acceptable increase in statistical uncertainties due to the co-magnetometer correction (see Sec.\ \ref{sec:MagReq}) of 5\% the laser based \magHg magnetometer performance is good enough for an eight fold improvement in nEDM sensitivity.
Given that future nEDM experiments will feature two independent precession volumes (and thus two \magHg co-magnetometers) the total improvement in nEDM sensitivity is further increased by a factor $\sqrt{2}$.
Taking this into account, the demonstrated signal quality is sufficient for an eleven fold improvement of statistical nEDM sensitivity compared to the performance of the PSI nEDM experiment in the 2016 data run of $\delta \dn \approx \SI{1E-25}{\ecm}$ per day.
We conclude that laser readout will be a necessary component for nEDM experiments using Hg co-magnetometers with total sensitivities in the low \SI{E-27}{\ecm} range.

Past neutron EDM experiments like the one performed at ILL\cite{Baker2006, Pendlebury2015PRD} or our own effort at PSI relied on spectral lamps while future experiments will all be based on lasers.
The derived signal model permits the analysis of lamp- and laser-based data in a common and consistent framework. 
This is particularly useful in studies that combine data sets from different experiments like the search for axion-like particles \cite{Abel2017} which in the future will probably combine data sets with Mercury magnetometers using spectral lamps and lasers.

\section*{\label{sec:Ack}Acknowledgements}
We acknowledge the great support of the PSI technical staff, specifically M.~Meier and F.~Burri. This work is part of the Ph.D. thesis of M.~F.~\cite{Fertl2013}.

\section*{Funding}
Swiss National Science Foundation (SNSF) (126562, 144473, 149211, 157079, 162574, 172639);
French Agence Nationale de la Recherche (ANR) (ANR-09-BLAN-0046, ANR-14-CE33-0007);
National Science Centre, Poland  2015/18/M/ST2/00056).





\end{document}